\PassOptionsToPackage{a4paper,margin=25mm,centering,asymmetric}{geometry}

\documentclass[pdflatex,sn-basic,oneside]{sn-jnl}

\setcitestyle{aysep={,}}

\usepackage[T1]{fontenc}

\usepackage{bm}
\usepackage{amsmath,amsfonts,amssymb,amscd,amsthm}
\usepackage[usenames,dvipsnames,svgnames]{xcolor}
\usepackage{graphicx}
\graphicspath{{../plots/paper/}{figures/}}

\usepackage{caption}
\usepackage{subcaption}
\usepackage{doi}
\usepackage{makecell}
\usepackage{hyperref}
\usepackage{acronym}
\usepackage{listings}
\usepackage{textgreek}
\usepackage{url}
\usepackage{color}
\usepackage{framed}
\usepackage{verbatim}
\usepackage{fancyvrb}
\usepackage{microtype} 
\usepackage[final]{pdfpages}
\usepackage{tikz}
\usepackage{booktabs}
\usepackage{multirow}
\usepackage{algorithm}
\usepackage{algpseudocode}

\usepackage[normalem]{ulem}

\usepackage{textcomp}
\usepackage{cancel}
\usepackage{mathrsfs}
\usepackage{stmaryrd}
\usepackage{parskip} 
\usepackage{import}
\usepackage{xifthen}
\usepackage{pdfpages}
\usepackage{hyperref}
\usepackage{float}
\usepackage{graphicx}
\usepackage{siunitx}
\usepackage{placeins}
\usepackage[labelfont=bf,skip=10pt]{caption}
\usepackage{enumitem}
\usepackage{cleveref}
\usepackage{import}
\usepackage{transparent}
\usepackage{xcolor}
\usepackage{longtable}
\catcode`\|=12\relax
\usepackage{pgfplots}
\pgfplotsset{compat=newest}
\usepackage{program}
\usepackage{tikzscale}
\usepackage{soul}
\usepackage{pmboxdraw}

\tikzset{
every picture/.style={
line width = 0.3mm} 
}
\pgfplotsset{
every axis/.append style={
line width = 0.3mm,
grid style={
    line width = 0.3mm,
},
tick style={
    line width = 0.3mm,
},
},
}

\makeatletter
\newsavebox{\tabcapsinglebox}
\long\def\@tablecaption#1#2{%
  \sbox\tabcapsinglebox{\tablecaptionfont\textbf{#1:~#2}}%
  \ifdim\wd\tabcapsinglebox<\textwidth
    \setbox\tabcapbox\vbox{\tablecaptionfont\noindent\usebox\tabcapsinglebox\vphantom{y}\par}%
  \else
    \setbox\tabcapbox\vbox{\tablecaptionfont\raggedright\noindent\textbf{#1: }#2\vphantom{y}\par}%
  \fi
  \box\tabcapbox%
}
\makeatother

\usepackage{graphicx,array}
\hypersetup{
breaklinks=true,
bookmarksopen=true,
pdftitle={Template article},    
pdfauthor={BadiaLab},     
colorlinks=true,       
linkcolor=black,          
citecolor=blue,        
filecolor=black,      
urlcolor=blue           
}
\definecolor{bg}{rgb}{0.93,0.93,0.93}

\acrodef{ode}[ODE]{ordinary differential equation}
\acrodef{pde}[PDE]{partial differential equation}
\acrodef{fe}[FE]{finite element}
\acrodef{fem}[FEM]{finite element method}
\acrodef{fcm}[FCM]{finite cell method}
\acrodef{DOF}[DOF]{degree of freedom}
\acrodefplural{DOF}[DOFs]{degrees of freedom}
\acrodef{agfem}[AgFEM]{aggregated finite element method}
\acrodef{cutfem}[CutFEM]{cut finite element method}
\acrodef{dg}[DG]{discontinuous Galerkin}
\acrodef{ls}[LS]{level set}
\acrodef{to}[TO]{topology optimisation}
\acrodef{nn}[NN]{neural network}
\acrodef{ldf}[LDF]{linear driving force}
\acrodef{dac}[DAC]{direct air capture}
\acrodef{mof}[MOF]{metal organic framework}
\acrodef{tvsa}[TVSA]{temperature vacuum swing adsorption}
\acrodef{SIMP}[SIMP]{solid isotropic microstructure with penalization for intermediate densities} 
\acrodef{ipcc}[IPCC]{Intergovernmental Panel on Climate Change}
\acrodef{sipg}[SIPG]{symmetric interior penalty Galerkin}
\acrodef{supg}[SUPG]{streamline upwind Petrov-Galerkin}
\acrodef{eit}[EIT]{electrical impedance tomography}

\newcommand{\tnor}[1]{{\left\vert\kern-0.25ex\left\vert\kern-0.25ex\left\vert #1 
\right\vert\kern-0.25ex\right\vert\kern-0.25ex\right\vert}}

\definecolor{mypurple}{HTML}{7E398D}

\begin{document}

\newcommand{\incfig}[2][1]{%
  \includegraphics[width=#1\linewidth]{#2.pdf}%
}

\newcommand{\mycomment}[1]{}

\newcommand{\updatedinfirstround}[1]{{#1}}
\newcommand{\removedinfirstround}[1]{}

\newcommand{\updated}[1]{{#1}}
\newcommand{\removed}[1]{}

\newcommand{\hsft}{11}

\makeatletter
\@twosidefalse
\@mparswitchfalse
\setlength{\evensidemargin}{\oddsidemargin}%
\makeatother

\title[Topology optimisation and \updated{simulated} state reconstruction for self-sensing grippers]
{Topology optimisation and \updated{simulated }state reconstruction for self-sensing \updatedinfirstround{piezoresistive}\removedinfirstround{soft} grippers
}


\author*[1]{\fnm{Connor N} \sur{Mallon}}\email{\textcolor{black}{Connor N Mallon: }connor.mallon@qut.edu.au}
\author[1]{\fnm{Zachary J} \sur{Wegert}}
\author[1]{\fnm{Anthony P} \sur{Roberts}}
\author[2]{\fnm{Joshua} \sur{Pinskier}}
\author[2]{\fnm{Harry} \sur{Bowman}}
\author*[1]{\fnm{Vivien J} \sur{Challis}}\email{\textcolor{black}{Vivien J Challis: }vivien.challis@qut.edu.au}

\affil*[1]{\orgdiv{School of Mathematical Sciences}, \orgname{Queensland University of Technology}, \orgaddress{\city{Brisbane}, \state{QLD}, \postcode{4000}, \country{Australia}}}

\affil[2]{\orgdiv{Data61 Robotics}, \orgname{CSIRO}, \orgaddress{\city{Pullenvale}, \state{QLD}, \postcode{4069}, \country{Australia}}}



\abstract{
\updatedinfirstround{We consider the }\updated{computational }\updatedinfirstround{design of a self-sensing robotic gripper that has the potential to reduce reliance on traditional external sensors.
We }\updated{computationally }\updatedinfirstround{investigate }\updated{the potential of }\updatedinfirstround{a novel sensing method that exploits the coupled electromechanical properties of piezoresistive materials to enable mechanical perception using sparse electrical measurements.}
\updatedinfirstround{In this theoretical proof-of-concept, we show that under }\updated{the idealised assumptions of }\updatedinfirstround{linear constitutive relationships and quasi-static gripper operation, }\updated{it is possible to}\updatedinfirstround{ reconstruct \updated{simple} contact shapes and gripper displacement fields using }\updated{a small number of simulated}\updatedinfirstround{
electrical measurements}\updated{, suggesting the potential of the proposed sensing technique.}
\updatedinfirstround{We use topology optimisation to promote the quality of displacement state recovery by adding an electrical performance objective to the mechanical compliance requirement of the
gripper.
Using the proposed reconstruction technique, we showcase the improved reconstruction ability of an electrically optimised gripper design through an increased robustness to noisy synthetic measurements compared to a gripper designed with only a mechanical objective in the topology optimisation problem.
In the future, detailed inclusion of non-linear effects and experimental realisation of this theoretical }\updated{proof-of-concept may }\updatedinfirstround{help remove the constraints of traditional sensors for robotic handling, }\updated{potentially facilitating sensing in settings where current sensing systems are impractical.}
}


\maketitle


\section{Introduction}


Automating tasks traditionally requiring manual labour is a cornerstone of modern industrial practice. Robotic systems have been developed to address a subset of these tasks that involve interaction with the physical environment. 
In particular, for applications requiring the handling of complex, delicate, and deformable objects, soft robotic grippers are especially well suited \citep{Shintake2018, Kim2013}. This follows from their inherent material compliance, enabling them to conform to the geometries of objects and apply gentle, evenly distributed forces \citep{Shintake2018, Rus2015}. \removedinfirstround{In this work, we consider aspects of soft gripper design to improve their practical performance.}


During gripper operation, measurements of contact forces and internal stresses are essential for ensuring effective grasps and preventing structural failure \citep{Hegde2023}. A structural model of a gripper can be developed to predict these quantities based on known boundary conditions. In practice, however, when the gripper interacts with an unknown object the boundary conditions at the contact interface are not known. A sensing mechanism is therefore required to obtain this boundary information.
Rigid sensors in the contact zone compromise the compliance of the gripper, involve complex wiring and are prone to damage \citep{VanDuong2021,Shimonomura2019}.
An alternative is to obtain missing information through a complex 3D vision system, although this may involve a cumbersome setup and be prone to occlusion \citep{Greenland2025,Zhan2025,VanDuong2021, Liu2024}. 
We instead consider the possibility of obtaining such information by utilising a \removedinfirstround{soft }piezoresistive material for the gripper and exploiting the piezoresistive effect, \updated{to facilitate sensing in cases where a vision system is not feasible.}
The piezoresistive effect describes the change in electrical resistivity of a material when subjected to mechanical stress \citep{Smith1954}. 
Current systems that integrate piezoresistive materials into grippers provide a low-dimensional electrical signal that can be used only to obtain low-order information, such as contact detection or object identification \citep{Stano2022,Shih2019,Georgopoulou2021,Georgopoulou2022}.
In contrast, we propose a new approach that enables recovery of an approximation of the entire displacement field and therefore the internal stresses and contact forces of the gripper based on electrical readings at a sparse set of convenient measurement locations. 

\updated{To the best of our knowledge, no prior work has attempted to recover the full displacement field of a gripper from sparse electrical measurements.
We consider this computational proof-of-concept as a first step investigating the potential of piezoresistive materials to provide self-sensing capability for robotic grippers.
As such, }\updatedinfirstround{we consider the simplest setting that still allows us to }\updated{investigate}\updatedinfirstround{ the potential of the sensing method.
Specifically, we assume that: the piezoresistive material lies at the relatively stiff end of the soft-material spectrum, for example, it is manufactured using a thermoplastic filament; the gripper does not undergo large deformations; and the gripper operates quasi-statically. These assumptions restrict the direct applicability of the models used in this work to a limited class of gripper applications, but they provide a foundation for future work addressing more general nonlinear behaviour.}
\updated{We further note that experimental validation of the piezoresistive electrical response and sensing methodology is outside of the scope of the current work. Recent works involving physical experiments on the piezoresistive response of printed materials that motivated this work can be found in the literature \cite{Georgopoulou2022,Georgopoulou2021,Georgopoulou2020,Riddervold2024}. We note that the constitutive behaviour of some of these filaments is linear in the reported experiments, which is consistent with the assumptions made in this work.
}

For displacement-field recoverability, the piezoresistive response must be measurable and information-rich. To provide this, we employ a topology optimisation approach to design the gripper's geometry, simultaneously maximising the piezoresistive response and mechanical performance to ensure reliable information recovery and high-quality grasps.
\updatedinfirstround{The mechanical performance of a gripper can be accounted for by a variety of different objective functions in the topology optimisation problem, depending on the application. A common approach is to consider the gripper as a compliant mechanism and to maximise the output displacement at the gripper tip for a given input force. This can either be done directly by maximising the output displacement \citep{Sun2022,Pinskier2024,Xie2024}, or by maximising the mutual strain energy \citep{Liu2018,Liu2022}. These objectives are considered to improve the gripper's ability to close around objects \citep{Pinskier2024}. 
An alterative objective is to measure how close the deformation of the gripper is to a target deformation, which can be used to design grippers that achieve specific motion \citep{Chen2021,Liu2021}.
Depending on the formulation and application, a minimisation of overall strain energy may also be considered as an additional mechanical objective, which can either act as a regularisation term for optimisation or be an important consideration when overall stiffness is desirable \citep{Liu2018,Pinskier2024}.
Topology optimisation approaches for more general compliant mechanisms are also extensively studied in the literature, with a variety of combinations of compliance and stiffness objectives \citep{Zhu2020}.}
Other works combine the compliance objective with additional objectives and constraints, such as for the stress \citep{DeLeon2015} or fundamental frequency \citep{Wang2024}.
To date, however, no work has combined a compliance objective with a piezoresistive response maximisation.
Work in topology optimisation that involves piezoresistive materials generally focuses on a single sensor design \citep{Zhu2016,Zhuang2023,Guo2024,Rubio2007,Giusti2014,Mello2012}. 
Recently, the design of structures with self-sensing as an additional characteristic have been investigated by \citet{Seifert2019,Seifert2019-2}, who use a functionally graded material to balance stiffness and sensing objectives. 
Consistent with the previous works focused on sensor design, we seek to maximise the piezoresistive response to improve the signal-to-noise ratio of the electrical measurements. For the first time we combine this with a compliant mechanism objective. Moreover, whereas existing designs target a single overall electrical response, we instead maximise signals across multiple electrode source-sink patterns and measurement locations.


Prior formulations for piezoresistive topology optimisation use density-based approaches that introduce interpolated materials into the topology optimisation problem \citep{Zhu2016,Zhuang2023,Guo2024,Rubio2007,Giusti2014,Mello2012,Seifert2019,Seifert2019-2}. Although convenient, these topology optimisation methods may induce non-physical features when considering the piezoresistive topology optimisation problem due to the presence of the interpolated material. To avoid this, we adopt an unfitted topology optimisation approach that eliminates interpolated materials altogether. Without re-meshing, unfitted finite element methods \citep{Badia2018,Burman2015} integrate the governing equations over the physical domain only. These methods are advantageous for topology optimisation in multi-physics contexts where the introduction of artificial densities are undesirable \citep{Jenkins2015,Villanueva2017}. Unfitted topology optimisation has been successfully applied to topology optimisation problems in various multi-physics contexts, including fluid-structure interaction \citep{Jenkins2015} and fluid-transport \citep{Villanueva2017}. To date, however, these methods have not been considered for electromechanical formulations such as the piezoresistive problem. In this work, we formulate an unfitted framework for the piezoresistive problem with a mixed mechanical and electrical objective that eliminates the introduction of artificial interpolated materials that can lead to non-physical features when considering the multi-physics formulation.


In the second part of this work, we propose a method for reconstructing unknown contact geometries based on sparse electrical measurements taken during gripper operation. In this work, we consider \emph{simulated} electrical measurements obtained when grasping a target object. Discarding all information about the target object except for these measurements, we attempt to reconstruct the shape and location of the contact object by solving an inverse problem.
With a reconstruction of a contact geometry, we can determine an approximation of the entire displacement field of the gripper by solving the forward problem. A related problem that has been extensively studied in the medical imaging literature is the classical problem of \ac{eit} that involves the reconstruction of unknown conductivity fields from electrical measurements \citep{youssef2024past}.
In recent decades, numerous studies have introduced force sensor skins based on tomographic reconstruction principles, allowing the use of soft materials to recover force locations without wiring in the contact region
\citep{Dai2019,Nagakubo2007,Alirezaei2007,Wu2022,Soleimani2022}.

\updated{
The sensor skin problem involves recovering the location of an unknown force along a surface.
In this work, we consider a related but distinct problem: recovering the location and shape of an unknown contact object.
The difference is that we require a contact formulation, i.e., we need to recover how deformed the gripper is when it comes into contact with the object, as well as where along the gripper's surface the contact occurs.
To the best of our knowledge, there is no existing work that solves this specific inverse problem by coupling a contact mechanics model with an electrical forward model to reconstruct unknown contact geometries from electrical measurements.
The closest examples of coupling mechanics into an electrical inverse problem are found in the self-sensing composites literature. \citet{Tallman2017} invert electrical impedance tomography conductivity data through an analytical piezoresistivity model to recover the displacement field of a specimen under known loading conditions.
Conversely, \citet{Hassan2023} reconstruct an a priori unknown damage geometry, such as through-holes and delaminations, from electrical impedance tomography data, but do so using a purely electrical forward model without coupling any mechanical equilibrium equation.
We consider a level set-based representation to be a natural choice here for parametrisation of the contact shape, and consider this level-set based reconstruction method as an extension of the current tomographic reconstruction literature to problems involving contact.}

The article is organised as follows. In Section \ref{sec:model-equations} we introduce the governing equations of the problem and in Section \ref{sec:numerical-discretisation} we present the numerical discretisation. In Section \ref{sec:topology-optimisation} we detail the proposed topology optimisation framework and obtain a new gripper design considering both electrical and mechanical objectives. In Section \ref{sec:state-recovery} we introduce the inverse framework for reconstructing unknown contact geometries and gripper displacement fields and demonstrate the enhanced performance of the new gripper design using synthetic data. 
Finally, Section \ref{sec:discussion}
discusses limitations of the theoretical proof-of-concept presented in the paper, and concluding remarks are given in \ref{sec:conclusion}.

\begin{figure}
	\centering
  	\incfig[0.8]{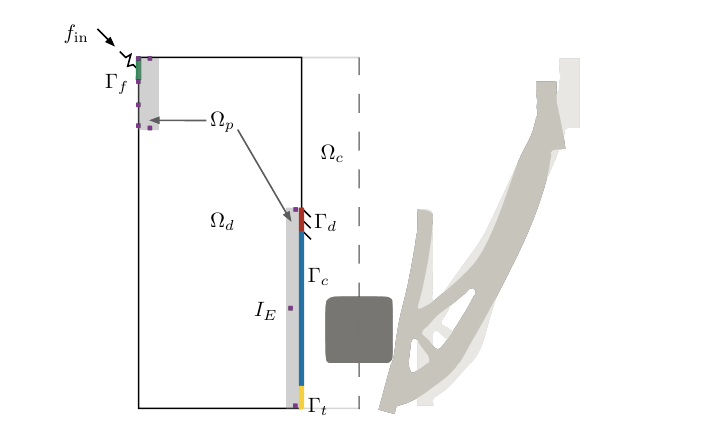}
		\caption{ 
    \updatedinfirstround{A depiction of the various integration domains}. 
		The model setup is composed of a passive (fixed solid) domain $\Omega_p$, shown in grey, a design domain $\Omega_d$ and a domain $\Omega_c$ containing the contact object.
		An input force $f_\mathrm{in}$ is applied at $45^\circ$ through a spring (representing the actuator stiffness) on the surface $\Gamma_f$ in green and a Dirichlet condition is prescribed on the displacement on the surface $\Gamma_d$ in red. The possible contact surface is denoted as the union of the surfaces $\Gamma_c$ in blue and $\Gamma_t$ in yellow. The gripper tip $\Gamma_t$ is used only in formulation of the topology optimisation problem. The measurement electrode locations are depicted by the set of purple points $I_E$.}
	\label{fig:design-domain-sym}
\end{figure}

\begin{figure}
	\centering
  \incfig[0.8]{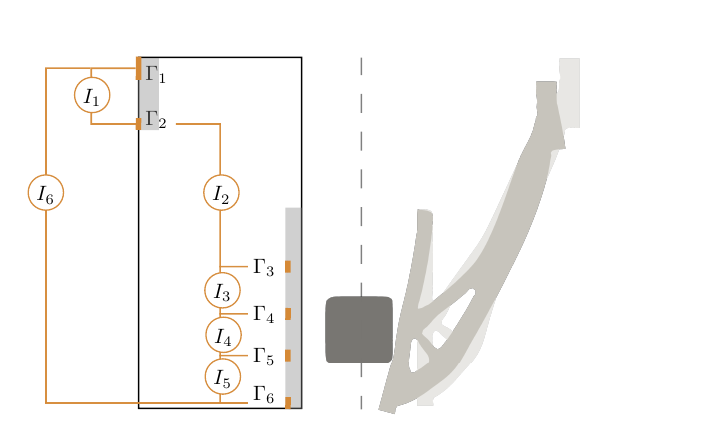}
	\caption{
		Superimposed current source-sink patterns. A current source-sink pattern \updatedinfirstround{$I_i$} is applied by setting a source electrode $\Gamma_i$ and a sink electrode $\Gamma_{i+1}$. The source-sink patterns $I_i$ for $i\in \{1,\ldots,6\}$ are overlaid here.
	}
	\label{fig:current-injection}
\end{figure}

\begin{figure}
	\centering
  \incfig[0.8]{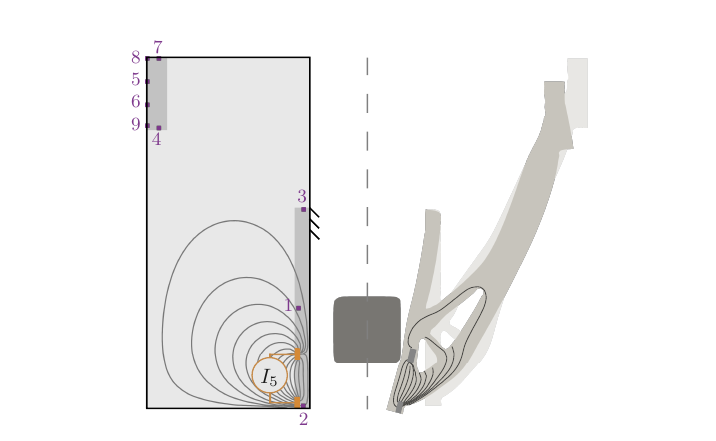}
	\caption{
    Illustrative example of a single current source-sink pattern $I_5$ \updated{for a completely solid gripper (left-hand side) and an example optimised gripper (right-hand side)}. Example current lines are shown flowing from the source electrode, $\Gamma_5$, to the sink electrode $\Gamma_6$. The numbered measurement electrode set $I_E$ is also shown in purple.
  }
	\label{fig:current-example}
\end{figure}

\section{Model equations}\label{sec:model-equations}
In this section we describe the continuum model for the coupled mechanical and electrical problems for a piezoresistive gripper. \updatedinfirstround{
The design domain adopted in this work is based on the one introduced by \cite{Liu2018} and is shown in Figure \ref{fig:design-domain-sym}. The left hand side of the symmetrical problem is considered, with a gripper visualised on the right to aid in interpretation of the diagram.}
We define $\Omega_{\text{grip}}$ as the domain occupied by the solid gripper material. This is composed of the passive domain $\Omega_p$ and a portion of the design domain $\Omega_d$ depending on the design. A formal definition for $\Omega_{\text{grip}}$ is given in Section \ref{sec:numerical-discretisation}.
Assuming linear elasticity and a conservation of charge, and noting that the piezoresistive effect couples the displacement and electric fields through a displacement-dependent conductivity \citep{Smith1954}, we obtain the following strong form of the governing equations for the interior of the domain:
\begin{equation}
\left\lbrace
\begin{aligned}
- \bm{\nabla} \cdot \bm\sigma(\bm{u}) &= \bm{0} &&\text{in}\  \Omega_\text{grip}, \\
- \bm{\nabla} \cdot (\kappa(\bm{u})\cdot\bm{\nabla}{\phi}) &= 0 &&\text{in}\ \Omega_\text{grip}, 
\end{aligned}
\right.
\end{equation}\label{eq:model-equations}where $ \bm{\sigma}(\bm{u}) = \lambda \text{tr}(\bm{\varepsilon}(\bm{u}))\bm{I}+2\mu \bm{\varepsilon}(\bm{u}) $ is the stress tensor, $\bm{\varepsilon}(\bm{u}) = \frac{1}{2}(\boldsymbol{\nabla}\boldsymbol{u} + (\boldsymbol{\nabla}{\boldsymbol{u}})^\top)$ is the \updatedinfirstround{small-strain symmetric tensor}, $\lambda$ and $\mu$ are the Lam\'e parameters given by $\lambda = (E\nu)/((1+\nu)(1-2\nu)) $ and $ \mu=E/(2(1+\nu))$ and $E$ and $\nu$ are the Young's modulus and Poisson's ratio, respectively, considering plane strain conditions. 

The displacement-dependent conductivity tensor $\kappa(\bm{u})$ for the 2D problem is defined as:
\updatedinfirstround{\begin{equation}
\kappa(\bm{u})=
\begin{bmatrix}
\kappa_{xx}(\bm{u}) & \kappa_{xy}(\bm{u})\\
\kappa_{xy}(\bm{u}) & \kappa_{yy}(\bm{u})
\end{bmatrix},
\end{equation}
with components approximated through the linear constitutive relation used by \cite{Giusti2014}:
\begin{equation}
\begin{aligned}
\kappa_{xx}(\bm{u}) &= \kappa_0\left(1-\pi_{11}\sigma_{xx}(\bm{u})-\pi_{12}\sigma_{yy}(\bm{u})\right),\\
\kappa_{yy}(\bm{u}) &= \kappa_0\left(1-\pi_{12}\sigma_{xx}(\bm{u})-\pi_{11}\sigma_{yy}(\bm{u})\right),\\
\kappa_{xy}(\bm{u}) &= -\kappa_0\pi_{44}\sigma_{xy}(\bm{u}).
\end{aligned}
\end{equation}}
\noindent where $\kappa_0$ is the base conductivity of the material under zero displacement and $\pi_{11},\pi_{12}$ and $\pi_{44}$ are the piezoresistivity coefficients. 
A homogeneous Dirichlet condition is prescribed on the displacement:
\begin{equation}
\bm{u} = \bm{0} \quad \text{on}\ \updatedinfirstround{\Gamma_d},
\end{equation} 
and traction boundary conditions are applied on the remainder of the boundary:
\begin{equation}
\left\lbrace
	\begin{aligned}
	\bm{\sigma}(\bm{u})\cdot\bm{n} &=
	\bm{f} - k_s\bm{u} &&\text{on}\ \Gamma_f, \\
	\bm{\sigma}(\bm{u})\cdot\bm{n} &= - \frac{1}{\epsilon} \langle -g \rangle_+ \bm{n}  &&\text{on}\ \Gamma_c, \\
	\bm{\sigma}(\bm{u})\cdot\bm{n} &= \bm{0} &&\text{on}\ \partial\Omega_\text{grip}\setminus(\Gamma_f\cup\Gamma_c\cup\Gamma_D),
	\end{aligned}
\right.
\label{eq:d-neumann}
\end{equation}
where $\bm{n}$ is the outward unit normal vector to the boundary, $\boldsymbol{f}$ is the forcing term with a magnitude $f_{\mathrm{in}}$ applied at $45^{\circ}$ on the input port $\Gamma_f$ as shown in Figure \ref{fig:design-domain-sym} and $k_s$ is the stiffness of the actuator at the input port. The condition on the contact surface $\Gamma_c$ follows a frictionless penalty method \citep[Eq. 4.73,][]{Yastrebov2013}, where $\epsilon$ is a small parameter, $\langle\ \rangle_+$ is the positive part function and $g$ is the gap function, defined as the distance between the gripper and the contact object. Using this method, a large normal reaction force is applied whenever the gripper intrudes into the object, i.e. when $g<0$. The remainder of the boundary is traction-free.




Along with the electrodes used to measure the electrical potential, we also must include electrodes to apply current to the system. This is done by applying Dirichlet conditions on the potential at some source-sink electrodes. 
We make clear here that these additional electrodes, which act as sink and source boundary conditions for the potential, are distinct from the measurement electrodes that are used only to measure the potential and do not affect the governing equations.
We select a few different locations for these source-sink electrodes and solve the coupled problem for each of them.
Using multiple current source-sink patterns increases the amount of independent information obtained, improving reconstruction accuracy \citep{Isaacson1986}. We adopt adjacent source-sink patterns, common in \ac{eit} \citep{Russo2017} which involves selecting sets of two adjacent electrodes as the source and sink electrodes. The source-sink patterns $I_i$ for $i\in \{1,\ldots,6\}$ are seen in Figure \ref{fig:current-injection} and 
involve the prescription of the following Dirichlet conditions on the potential:
\begin{equation}
\phi = 1 \ \text{on}\ \Gamma_i, \quad \phi
= 0 \ \text{on}\ \Gamma_{i+1}, \quad i=1,\ldots,6,
\end{equation}
where $\Gamma_7\equiv\Gamma_1$. An example of one such current source-sink pattern is shown in Figure \ref{fig:current-example}.
On the remainder of the boundary, the electrical insulation condition is applied:
\begin{equation}
(\kappa(\bm{u})\cdot\bm{\nabla}{\phi
})\cdot\bm{n} = 0 \ \text{on}\ \partial\Omega_\text{grip}\setminus(\Gamma_i\cup\Gamma_{i+1}).
\end{equation}

The values of the physical constants introduced in this section are given in Table \ref{tab:physical-constants}. \updatedinfirstround{The material is assumed to have the Young's modulus of a relatively stiff thermoplastic (as per the gripper printed using piezoresistive filament for the experiment in Section~\ref{sec:discussion}). 
As per the constitutive equations above, the material is assumed to behave linearly elastically and geometric nonlinearities are ignored. The validity of both of these assumptions is analysed in more detail in Section~\ref{sec:discussion}.
Quasi-static conditions are also assumed, implying sufficiently slow operation speeds. Specifically, time-dependent effects such as percolation, hysteresis and rate dependence are ignored. We also assume that signal drift is either calibrated for or controlled experimentally, so that drift does not affect electrical measurements. The above assumptions restrict the applicability of the current work. Extensions to model highly deformable elastomers and faster operation timescales, and to account for drift in a specific context are important future research directions. }

\begin{table}[h]
  \centering
  \caption{Physical constants}
  \label{tab:physical-constants}
  \begin{tabular}{l r l}
    \toprule
    Constant & {Value} & Unit \\
    \midrule
    Input force density, $f_{\mathrm{in}}$          & 141      & \unit{MPa}   \\ 
    Input actuator stiffness, $k_s$ & 10$\times 10^{10}$ & \unit{N\cdot m^{-1}} \\
    Elastic modulus, $E$                        & 11.8 & \unit{MPa}  \\
    Poisson's ratio, $\nu$                      & 0.45               & \\
    Base conductivity, $\kappa_0$        & 1               & \unit{S\cdot m^{-1}} \\
    Piezoresistive coefficient, $\pi_{1 1}$  & 5 $\times 10^{-7}$ & \unit{Pa^{-1}} \\ 
    Piezoresistive coefficient, $\pi_{1 2}$  & 5 $\times 10^{-7}$ & \unit{Pa^{-1}} \\ 
    Piezoresistive coefficient, $\pi_{4 4}$  & 5 $\times 10^{-7}$ & \unit{Pa^{-1}} \\ 
    \bottomrule
  \end{tabular}
\end{table}

\section{Numerical discretisation}\label{sec:numerical-discretisation}

\updatedinfirstround{
In this section, we describe the numerical discretisation of the problem defined in Section \ref{sec:model-equations}. 
For this particular problem, we note again that it is important to define a discretisation that allows for an exact representation of the geometry of the gripper.
If we were to adopt a density-based approximation of the geometry, as is common in topology optimisation, intermediate densities would be introduced to represent the geometry. For the piezoresistive optimisation problem, intermediate densities have the potential to be exploited by the optimisation algorithm to improve the piezoresistive response by allowing current to flow through unphysical regions. This is in contrast to the classic case of compliance minimisation, where it is easy to use appropriate penalisation to ensure that intermediate densities 
lead to a reduction in stiffness
and therefore are not exploited. For this reason, we adopt a CutFEM \citep{Burman2015} approach to define the geometry based on level set functions that allows for an exact representation of integration domains without the need for remeshing. We emphasise here that the particular unfitted topology optimisation approach is not a new contribution of this work and is described in detail by \cite{Wegert2025}. Instead, the novelty of this work lies in applying this approach to the piezoresistive design problem. We therefore recall the relevant details here and indicate how they apply in the present setting.
}

\begin{figure}[!htbp]
	\centering
    \includegraphics[width=0.4\linewidth]{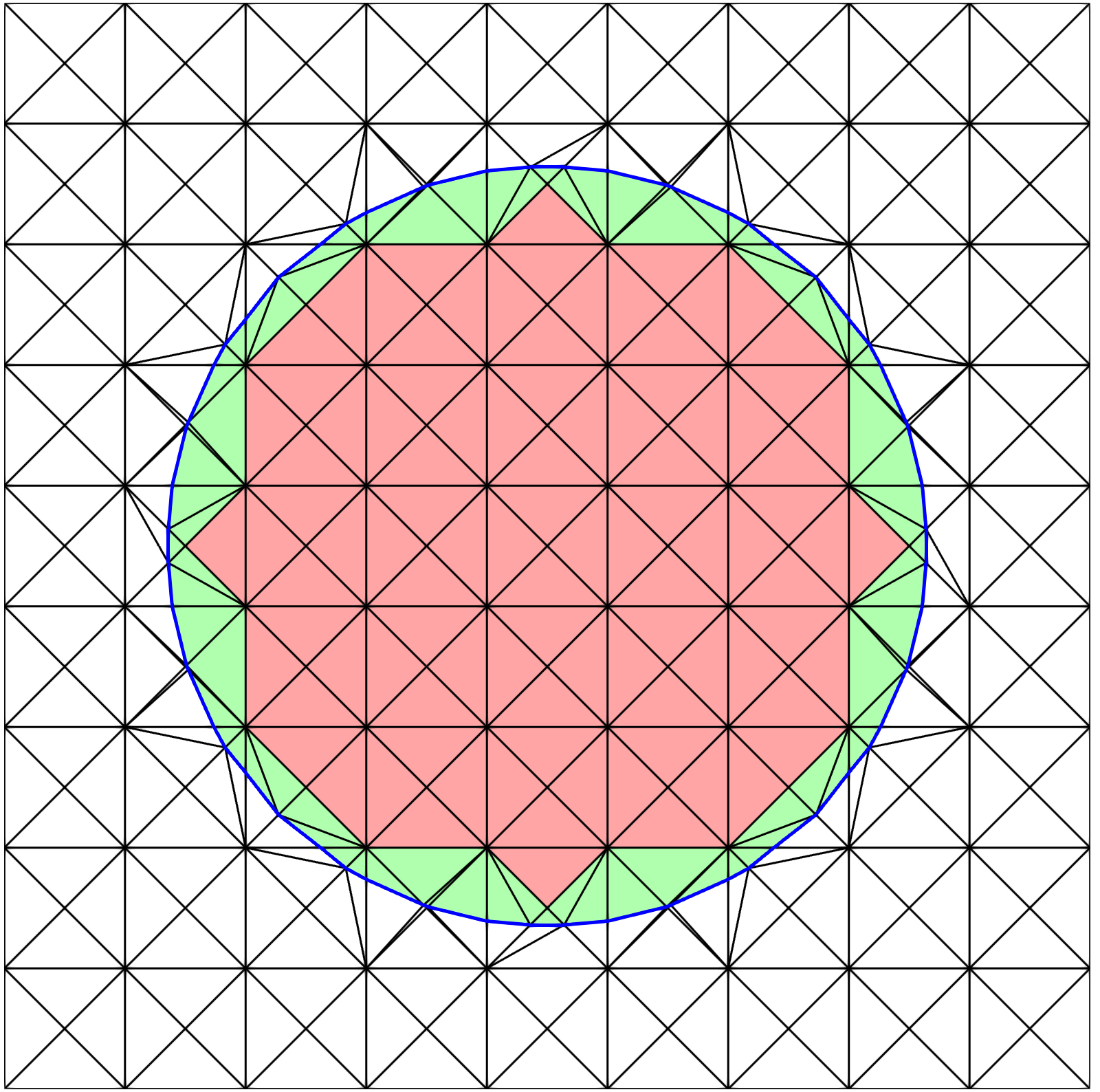}
	\caption{
    An example subtriangulation of a domain $\Omega$ defined by a level set $\varphi (x) \in V^*$. The green cells are those in $\Omega$ that include the interface $\Gamma$ and the red cells are the remaining internal
cells in $\Omega$. This figure is reproduced with permission from \cite{Wegert2025-1}.
    }
	\label{fig:subtrian}
\end{figure}

The CutFEM approach used to solve the problem defined in Section \ref{sec:model-equations} is now described. Let $\mathcal{T}^*_{h}$ represent a conforming, simplicial, unstructured partition (mesh) of the background domain $\Omega^*\doteq\Omega_d\cup\Omega_p\cup\Omega_c$ with a characteristic mesh size $h$. Firstly, we define a nodal Lagrangian finite element space of order 1 on $\mathcal{T}^*_h$ as:
\begin{equation}
  V^{*} \doteq \{ v_{} \in H^1(\Omega^*): v_{}\vert_{K} \in P_1(K) \ \forall\ K \in \mathcal{T}^*_h \},
\end{equation}
where $P_1$ is the space of polynomials with maximum degree 1 in each variable. Using a restricted background space $V_{\text{grip}} \doteq \{ v\vert_{\Omega_d \cup \Omega_p} : v \in V^* \}$,
we define a level set function $\varphi_{\text{grip}} \in V_{\text{grip}}$ to describe the (solid) gripper domain $\Omega_\text{grip}(\varphi_{\text{grip}}) = \{ x \in \Omega_{\text{grip}}: \varphi_\text{grip}(x) < 0 \}$ and boundary $\Gamma_\text{grip}(\varphi_{\text{grip}}) = \{ x \in \Omega_{\text{grip}}: \varphi_\text{grip}(x) = 0 \}$ and introduce a second level set function $\varphi_\text{obj} \in V^*$ to describe the geometry of the contact object.
\updatedinfirstround{For clarity, we now write the discretisation explicitly. Let $\{N_a^*\}_{a=1}^{n^*}$ denote the continuous nodal $P_1$ basis on the background mesh $\mathcal{T}_h^*$. The level set fields are approximated as}
\begin{equation}
\updatedinfirstround{\varphi_{\text{grip}}(\bm{x}) = \sum_{a=1}^{n^*} \widehat{\varphi}^{\text{grip}}_a\,N_a^*(\bm{x}),\qquad
\varphi_{\text{obj}}(\bm{x}) = \sum_{a=1}^{n^*} \widehat{\varphi}^{\text{obj}}_a\,N_a^*(\bm{x}).}
\end{equation}
\updatedinfirstround{Here, $\widehat{\varphi}^{\text{grip}}_a$ and $\widehat{\varphi}^{\text{obj}}_a$ are nodal degrees of freedom (equivalently nodal values at $\bm{x}_a^*$), and the vectors $\widehat{\boldsymbol\varphi}^{\text{grip}}$ and $\widehat{\boldsymbol\varphi}^{\text{obj}}$ collect these coefficients. The physical gripper domain $\Omega_\text{grip}(\varphi_{\text{grip},h})$ is then induced by $\varphi_{\text{grip},h}$ through the following approach.
Firstly, for each cell in the mesh, the level set function $\varphi_{\text{grip}}$ is used to determine whether the cell is in the interior of $\Omega_\text{grip}(\varphi_{\text{grip}})$, in the exterior, or cut by $\Gamma_\text{grip}(\varphi_{\text{grip}})$ by inspecting the signs of the level set values at the cell's nodes. The cells in the interior are those for which $\varphi_{\text{grip}}$ is negative at all nodes, while the cells in the exterior are those for which $\varphi_{\text{grip}}$ is positive at all nodes. The cells cut by $\Gamma_\text{grip}(\varphi_{\text{grip}})$ are those for which $\varphi_{\text{grip}}$ has both positive and negative values at its nodes. For the cells cut by $\Gamma_\text{grip}(\varphi_{\text{grip}})$, a subtriangulation is performed to define a piecewise linear approximation of the interface within the cell, and to determine the portion of the cell that lies in the interior and exterior of $\Omega_\mathrm{grip}(\phi_\mathrm{grip})$. After performing this procedure for all cells, a cut mesh, such as the one shown in Figure \ref{fig:subtrian}, is defined that fully describes the geometry of $\Omega_\text{grip}(\varphi_{\text{grip}})$ and $\Gamma_\text{grip}(\varphi_{\text{grip}})$. The Gauss points for integration in the cut cells are updated based on the piecewise linear interface location so that the integration domain is defined by the portion of the cell in the interior of $\Omega_\mathrm{grip}(\phi_\mathrm{grip})$.}

\updatedinfirstround{With this implementation of the geometry, we are able to define restricted finite element spaces for the solution fields. The scalar finite element space for the approximation of the electrical potential field is given by:
\begin{equation}
  V \doteq \{ v_{} \in H^1(\Omega_\text{grip}(\varphi_{\text{grip}})): v_{}\vert_{K} \in P_1(K) \ \forall\ K \in \mathcal{T}_h \},
\end{equation}
where $\mathcal{T}_h\subset\mathcal{T}^*_h$ is the cut mesh composed of the union of the cells in the interior of $\Omega_\text{grip}(\varphi_{\text{grip}})$ and the cells cut by $\Gamma_\text{grip}(\varphi_{\text{grip}})$. 
The cut mesh is obtained by collecting the cells that are tagged, in the procedure described above, as being in the interior or cut by $\Gamma_\text{grip}(\varphi_{\text{grip}})$. Note that this cut mesh includes the entire cut cells and not just the portion of the cut cells in the interior. 
Similarly, the vector finite element space for approximation of the displacement field is defined as:
\begin{equation}
  \mathbf{V} \doteq \{ v_{} \in [H^1(\Omega_\text{grip}(\varphi_{\text{grip}}))]^d: \bm{v}_{}\vert_{K} \in [P_1(K)]^d \ \forall\ K \in \mathcal{T}_h \}.
\end{equation}
Again, for clarity, we write out the discretisation explicitly.} 
\updatedinfirstround{Let $\{N_a\}_{a=1}^{n_\phi}$ denote the continuous nodal $P_1$ basis for the scalar space $V$ on $\mathcal{T}_h$, and let $\{\bm{N}_a\}_{a=1}^{n_u}$ denote the associated vector-valued $P_1$ basis for $\mathbf{V}$. The state variables are approximated as}
\begin{equation}
\updatedinfirstround{\bm{u}(\bm{x}) = \sum_{a=1}^{n_u} U_a\,\bm{N}_a(\bm{x}), \qquad
\phi_{i}(\bm{x}) = \sum_{a=1}^{n_\phi} \Phi_{i,a}\,N_a(\bm{x}), \quad i\in I_A.}\label{eq:discrete-solutions}
\end{equation}
\updatedinfirstround{
Here, $\bm{u}$ and $\phi_{i}$ are the displacement and potential fields, $U_a$ and $\Phi_{i,a}$ are the corresponding nodal coefficients (degrees of freedom), $n_u$ and $n_\phi$ are the numbers of displacement and potential degrees of freedom, and $i$ indexes the source-sink patterns in $I_A$ (as shown in Figure~\ref{fig:current-injection}). In summary, the solution fields are represented as linear combinations of piecewise linear basis functions on the cut mesh $\mathcal{T}_h$, with nodal coefficients as degrees of freedom where the mesh $\mathcal{T}_h$ is induced by the level set function $\varphi_{\text{grip}}$ describing the gripper geometry.}

The discretised weak formulation of the problem defined in Section \ref{sec:model-equations} can now be described. \updatedinfirstround{Note here that the integration domains of the integrals in the weak formulation are defined by the level set functions $\varphi_{\text{grip}}$ and $\varphi_{\text{obj}}$ as described above.}
We aim to find a solution 
$\bm{u} \in \mathbf U_{0}$ and $
\phi_i \in U_{i}$ for all $i \in I_A$
such that:
\begin{equation}
R_d(\bm{u},\bm{v},\varphi_{\text{grip}},\varphi_{\text{obj}}) + \sum_{i\in I_A} R_\phi((\bm{u}_{},\phi_{i}),(\bm{v}_{},\eta_{i}),\varphi_{\text{grip}})=0,\quad 
\forall\,\bm v\in \mathbf V_{0},\;
\forall\,\eta_i\in V_i,
\label{eq:weak-formulation}
\end{equation} where $\mathbf{U}_{0}$ and $\mathbf{V}_{0}$ are the set of functions in $\mathbf{V}_{}$ that are zero on $\Gamma_d$, $U_{i}$ is the set of functions in $V_{}$ that are 1 on $\Gamma_{i}$ and 0 on $\Gamma_{i+1}$ and $V_{i}$ is the set of functions in $V_{}$ that are 0 on both $\Gamma_{i}$ and $\Gamma_{i+1}$. 
The structural component of the residual $R_d$ is defined as:
\begin{equation}
  \begin{aligned}
    R_d(\bm{u}_{},\bm{v}_{},\varphi_{\text{grip}},\varphi_{\text{obj}}) &\doteq 
    \int_{\Omega_\text{grip}(\varphi_{\text{grip}})} \sigma(\bm{u}_{}):\varepsilon(\bm{v}_{}) \ \mathrm{d}\Omega 
    + \int_{\Gamma_c} \frac{1}{\epsilon} \langle \varphi_{\text{obj}} \vert_{\bm{x}+\bm{u}}\rangle_+ \bm{v} \cdot \bm{n} \ \mathrm{d}\Gamma \\
      &+ \int_{\Gamma_f} (k_s \bm{u}_{} -\bm{f})\cdot \bm{v}_{}\  \mathrm{d}\Gamma 
    +S_d(\bm{u}_{},\bm{v}_{},\varphi_{\text{grip}}),
  \end{aligned}
\label{eq:Rd}
\end{equation}

with a stabilisation term $S_d$ to be defined below. The signed distance function is used as an approximation of the gap function in Eq. \eqref{eq:d-neumann} when evaluated at $\bm{x}+\bm{u}$ for the Gauss point $\bm{x}$ on the contact surface.
The electrical component of the residual $R_\phi$ is then defined as:
\begin{equation}
  \begin{aligned}
    R_\phi((\bm{u}_{},\phi_{}),(\bm{v}_{},\eta_{}),\varphi_{\text{grip}}) &\doteq  
     \int_{\Omega_\text{grip}(\varphi_{\text{grip}})} \kappa(\bm{u}_{}) \nabla \phi_{} \cdot \nabla \eta_{} \  \mathrm{d}\Omega 
    +S_\phi(\phi_{},\eta_{},\varphi_{\text{grip}}),
  \end{aligned}
\label{eq:residual-phi}
\end{equation}
with a stabilisation term $S_\phi$. The stabilisation terms $S_d$ and $S_\phi$ are included to avoid ill-conditioning of the respective problems.
The stabilisation of the structural problem is given by: 
\begin{equation}
  \begin{aligned}
S_d(\bm{u},\bm{v}_{},\varphi_{\text{grip}}) 
\doteq 
\int_{\Omega_\text{grip}(\varphi_{\text{grip}})}\psi_{\bm{u}}\boldsymbol{u}\cdot\boldsymbol{v}~\mathrm{d}\Omega 
+\gamma_{\bm{u}}(\lambda+\mu)\sum_{F\in\mathscr{T}_G}\int_{F} h_F^3\llbracket\boldsymbol{n}_F\cdot\boldsymbol{\nabla u}\rrbracket\cdot\llbracket\boldsymbol{n}_F\cdot\boldsymbol{\nabla v}\rrbracket~\mathrm{d}F,
  \end{aligned}\label{eq:Sd}
\end{equation}
where $\psi_{\bm{u}}$ is an indicator function for volumes disconnected from $\Gamma_d$ (see \cite{Wegert2025} for implementation details), $\gamma_{\bm{u}}$ is a stabilisation parameter, $h_F$ is the characteristic face length, $\llbracket \cdot \rrbracket$ is the jump operator, $\boldsymbol{n}_F$ is the outward normal vector on the face $F$ and $\mathscr{T}_G$ is the set of all faces cut by $\Gamma_\text{grip}(\varphi_{\text{grip}})$.
\updatedinfirstround{One issue with the unfitted approach for topology optimisation is the potential for singularity of the stiffness matrix caused by isolated subdomains that form during the optimisation process that are disconnected from the Dirichlet boundary $\Gamma_d$.  This is due to the fact that these isolated subdomains contain displacement modes that are not fixed by the Dirichlet conditions.
To fix this issue, we define an indicator function $\psi_{\bm{u}}$ that identifies the isolated volumes and, using this indicator function, add a mass term that is activated in the isolated volumes to provide control over the displacement modes in these volumes. 
The indicator function is defined based on a connectivity analysis of the cut mesh $\mathcal{T}_h$ that identifies the volumes that are disconnected from $\Gamma_d$. 
The corresponding stabilisation term is the first term in Equation \eqref{eq:Sd}.}
\updatedinfirstround{
  Another issue that can arise using an unfitted discretisation is the presence of small cut cells that result in arbitrarily small entries in the stiffness matrix and thus ill-conditioning. 
The remedy to this issue is to add a consistent stabilisation term in the vicinity of the interface that provides control over the solution in the small cut cells.  
The particular term used here is a ghost penalty term described by \cite{Burman2015} that penalises the jump in the normal derivative across faces cut by the interface $\Gamma_\text{grip}(\varphi_{\text{grip}})$, which is the second term in Equation \eqref{eq:Sd}.}
The stabilisation of the electrical problem is defined similarly as:
\begin{equation}
  S_\phi((\bm{u}_{},\phi_{}),(\bm{v}_{},\eta_{}),\varphi_{\text{grip}}) 
  \doteq \int_{\Omega_\text{grip}(\varphi_{\text{grip}})}\psi_i\phi\eta~\mathrm{d}\Omega \\ 
  +\gamma_\phi k_0 \sum_{F\in\mathscr{T}_G}\int_{F} h_F\llbracket\boldsymbol{n}_F\cdot\boldsymbol{\nabla}\phi\rrbracket\cdot\llbracket\boldsymbol{n}_F\cdot\boldsymbol{\nabla }\eta\rrbracket~\mathrm{d}F,
\end{equation}
where $\psi_i$ is an indicator function for volumes disconnected from both
$\Gamma_{i}$ and $\Gamma_{i+1}$ and $\gamma_\phi$ is again a stabilisation parameter. The values of the numerical constants introduced in this section are given in Table \ref{tab:numerical-constants}. 

\updatedinfirstround{Finally, we derive the fully discrete system. Substitution of the expansions in Equation \eqref{eq:discrete-solutions} into Equations \eqref{eq:Rd} and \eqref{eq:residual-phi} yields the algebraic residual systems}
\begin{equation}
\updatedinfirstround{\mathbf{R}_u(\mathbf{U}) = \mathbf{0}, \qquad
\mathbf{R}_{\phi,i}(\mathbf{U},\boldsymbol{\Phi}_i) = \mathbf{K}_{\phi,i}(\mathbf{U})\,\boldsymbol{\Phi}_i-\mathbf{b}_i = \mathbf{0},\quad i\in I_A,}
\end{equation}
\updatedinfirstround{where $\mathbf{U}$ and $\boldsymbol{\Phi}_i$ collect the nodal coefficients of displacement and electrical potential, respectively. Here, $\mathbf{R}_u$ denotes the assembled nonlinear mechanical residual vector, $\mathbf{R}_{\phi,i}$ denotes the assembled electrical residual vector for source-sink pattern $i$, $\mathbf{K}_{\phi,i}(\mathbf{U})$ denotes the assembled stabilised deformation-dependent conductivity (stiffness) matrix, and $\mathbf{b}_i$ denotes the corresponding right-hand side vector for the electrical system. 
}
\updatedinfirstround{The coupled problem is solved in a staggered manner for each configuration. First, the nonlinear structural system $\mathbf{R}_u(\mathbf{U})=\mathbf{0}$ is solved with Newton's method. Then, with the converged displacement field fixed, the electrical systems $\mathbf{K}_{\phi,i}(\mathbf{U})\,\boldsymbol{\Phi}_i=\mathbf{b}_i$ are solved for each source-sink pattern $i\in I_A$.}

\begin{table}[h]
  \centering


  \captionsetup{labelformat=empty}
  \caption{\makebox[\linewidth][c]{\textbf{Table~\thetable:} \updatedinfirstround{Numerical constants}}}  

  \label{tab:numerical-constants}
  \begin{tabular}{l l }
    \toprule
    Constant & {Value}\\
    \midrule
    $\epsilon$                 & $10^{-11}$              \\
    $\gamma_{\phi}$              & 10               \\
    $\gamma_{\bm{u}}$              & $10^{-4}$           \\
    \bottomrule
  \end{tabular}
\end{table}


\section{Topology optimisation}\label{sec:topology-optimisation}

In the first part of this work we perform a level set-based topology optimisation to find an optimised gripper design considering both mechanical and electrical objectives. The gripper should ideally be optimised for performance on \textit{general} objects. As such, a choice must be made about the contact term (the second term in Eq. \eqref{eq:Rd}), which currently depends on the shape of a specific object $\varphi_\text{obj}$. A commonly adopted choice here is to instead use a simple spring term $\int_{\Gamma_t} k_t \bm{u} \cdot \bm{v} \ \mathrm{d}\Gamma$ at the gripper tip $\Gamma_t$ as the contact term.
We adopt this approach for the topology optimisation step 
and set the spring constant $k_t$ to $5\times 10^7$ \unit{N\cdot m^{-1}} and reduce the input force to $1.41N$ to recover a structural residual $R_{d,k_t}$ similar to that defined by \cite{Liu2018}. 

\updatedinfirstround{The setup geometry and boundary conditions are selected because of the practicality of experimental testing and availability of published optimised designs for this setup that show performance sensitivity to gripper geometry experimentally \citep{Xie2024,Liu2018,Liu2023}. The formulation of the problem is based on the work of \cite{Liu2018} because of the experimental validation performed in that work for the resulting designs that suggest validity of the formulation.} This structural residual is designed to avoid the emergence of artificial hinges during the topology optimisation. The topology optimisation problem can now be defined as follows:
  \begin{equation}
  \begin{aligned}
    \underset{\varphi \in V^1 }{\text{max}} \ \  \updatedinfirstround{J\doteq\ }  
    \alpha_S J_S + \alpha_E J_E \\
    \text{s.t}  \ R_{d,k_t}(\bm{u}_{},\bm{v}_{},\varphi) + \sum_{i\in I_A} R_\phi((\bm{u},\phi_{i}),(\bm{v}_{},\eta_{i}),\varphi) 
    = 0,& \quad \forall\,\bm v\in \mathbf V_{0},\;
\forall\,\eta_i\in V_i,\\
          \ \quad \quad \  \mathrm{Vol}(\varphi) = V_{\text{req}} ,&\\
  \end{aligned}
\label{eq:topopt-problem}
\end{equation}
where $\alpha_S$ and $\alpha_E$ are the objective weights, $\mathrm{Vol}(\varphi)$ is the volume of the gripper and $V_{\text{req}}$ is the required volume, taken to be 25\% of the volume of the background domain.
The first term of the objective function $J_S$ corresponds to maximising a \updatedinfirstround{tip-region displacement measure }as a proxy term for the compliance around general objects in order to ensure quality grasps \citep{Pinskier2024}:
\begin{equation}
    J_S = \int_{\Gamma_t} \bm{u}\  \mathrm{d}\Gamma.
\end{equation}
\updatedinfirstround{A boundary integral is used for the tip region instead of taking the value of a single degree of freedom at the tip because the latter may lead to designs that are more sensitive to small changes in the geometry.}
\updated{We choose this objective function because it is a smooth variation of the tip displacement objective that was seen to produce designs with experimentally verified mechanical performance \cite{Xie2024,Pinskier2024}. The design of the mechanical objective function is not the main contribution of this work and it is only used to ensure that the gripper has a reasonable mechanical performance. As such, we have not included a comparison with other mechanical objective functions because we consider that this would be outside the scope of the current work. However, we note that alternative formulations may produce designs with improved mechanical performance.}

The second term in the objective function is a maximisation of the electrical sensitivity, measured by the sum of the squared differences between the loaded and unloaded potentials evaluated at the measurement electrodes for each source-sink pattern:
\begin{equation}
    J_E = 
    \sum_{i =3}^5 
    \big\lVert 
    \phi_i - \phi_{i,0} 
    \big\rVert_{\ell^2(I_E)},
\end{equation}
where $\phi_{i,0}$ is the potential field for the unloaded state of the electrode arrangement $i$.
\updatedinfirstround{The difference between the loaded and unloaded potential is the signal that can be measured experimentally during gripper operation. By maximising the sum of this quantity over the electrodes, we get a measure of the strength of the overall signal. It is expected that by maximising this signal will improve the ability to recover information from the electrical measurements, particularly in the presence of noise.}

\updatedinfirstround{
The proposed electrical objective function $J_E$ is }\updated{selected}\updatedinfirstround{ to promote recovery of state information while keeping the topology optimisation and reconstruction problems uncoupled.
A more direct approach to promote recovery would be to include a reconstruction step in the topology optimisation loop and directly optimise for the reconstruction performance. Such an approach may further improve the recovery performance, however it would also lead to a significant increase in }\updated{computational }\updatedinfirstround{complexity of the topology optimisation problem} \updated{ by requiring solution of the reconstruction problem at each iteration of the topology optimisation algorithm. We consider this to be prohibitively expensive for the current work and it is not pursued
here.}
\updatedinfirstround{A key aim of the current work is to investigate the potential for a proxy term such as the proposed electrical objective $J_E$ to improve recovery performance without coupling the two problems.}
\updated{The idea of maximising the difference between the loaded and unloaded potentials at a particular location has been adopted previously in piezoresistive topology optimisation \citep{Rubio2007,Mello2012} and our work extends this idea to the current setting by considering multiple source-sink patterns and multiple measurement electrodes.}
\updatedinfirstround{An even simpler approach for the electrical objective would be to maximise the difference between the loaded and unloaded potentials integrated over the entire domain, as done in previous works on piezoresistive topology optimisation \citep[e.g.,][]{Giusti2014}.
However, that approach may lead to designs that only improve the signal strength away from the measurement electrodes} \updated{and is less effective when considering sensing at a set of known electrode locations. We consider that the proposed electrical objective function is a more appropriate choice for our application, because it directly targets the signal strength at the measurement electrodes. We show that such an approach can be used to improve the recovery of state information from simulated electrical measurements.}

We only include the source-sink patterns \updatedinfirstround{in $J_E$} that involve electrodes on the edge closest to the object here, that is $I_i$ for $i \in {3,4,5}$, which are assumed to be the most relevant for obtaining information on the contact edge. If all source-sink patterns are included, the optimisation may be dominated by less important patterns.


\updatedinfirstround{
In this paper we use the classical augmented Lagrangian method discussed by \cite{NocedalWright2006} to solve the constrained optimisation problem:
\begin{equation}\label{eqn: ALM opt prob}
\begin{aligned}
    \underset{\varphi\in V^1}{\min}~
      &\mathcal{L}(\varphi)\doteq J(\varphi)-\sum_{i=1}^{N}\left\{\lambda_iC_i(\varphi)+\frac{1}{2}\Lambda_iC_i(\varphi)^2\right\},
\end{aligned}
\end{equation}
where $\mathcal{L}$ is the augmented Lagrangian functional and $C_i(\varphi)$, $i = 1, \dots, N$, \updated{are equality constraint functions, each defined such that $C_i(\varphi) = 0$ when the corresponding constraint is satisfied. Depending on the application, $C_i$ may represent, for example, a volume constraint, a stress constraint \citep{Du2024}, or a manufacturing constraint \citep{Zhou2015}. In this work, we consider only a single ($N=1$) volume constraint,
\begin{equation}
    C_1(\varphi) = \mathrm{Vol}(\varphi) - V_{\text{req}},
\end{equation}
consistent with Eq.~\eqref{eq:topopt-problem}.}
In the above $\lambda_i$ and $\Lambda_i$ are the Lagrange multipliers and penalty parameters, respectively, that are updated in the standard way described by \cite{NocedalWright2006}. To minimise this Lagrangian functional, we update the level set $\varphi$ as per the evolution approach \citep{Burman2018}:
\begin{equation}\label{eqn: transport}
    \begin{cases}
        \dfrac{\partial \varphi(t,\boldsymbol{x})}{\partial t}+\boldsymbol{\beta}\cdot\boldsymbol{\nabla}\varphi(t,\boldsymbol{x})=0,\\
        \varphi(0,\boldsymbol{x})=\varphi_0(\boldsymbol{x}),\\
        \boldsymbol{x}\in D,~t\in(0,T),
    \end{cases}
\end{equation}
where $\boldsymbol{\beta}$ is a velocity field. In this work $\boldsymbol{\beta}$ is computed as $\boldsymbol{\beta}=\boldsymbol{n}g$, where $g$ is a regularised descent direction and $\boldsymbol{n}$ is the outwards facing normal to the boundary of the domain. 
The regularised descent direction is found by adopting a Hilbertian extension-regularisation framework \citep{ALLAIRE20211}. This involves solving the following problem: {Find $g\in V^1$ such that}
\begin{equation}\label{eqn: ALM ext shape sens}
    \langle g,w\rangle_H=d\mathcal{L}(\varphi;w)~\forall w\in V^1. 
\end{equation}
This determines a descent direction $g$ that improves the augmented Lagrangian $\mathcal{L}$ using the $H^1$ inner product:
\begin{equation}
    \langle u,v\rangle_{H}=\int_D\left(\alpha^2\boldsymbol{\nabla} u\cdot\boldsymbol{\nabla} v+uv\right)\mathrm{d}\boldsymbol{x},
\end{equation}
where $\alpha$ is the regularisation length scale \citep{Allaire2016}.
The directional derivative $d\mathcal{L}(\varphi;w)$ is computed using the adjoint method, with partial derivatives computed using automatic differentiation using the technique detailed in \cite{Wegert2025}. 
Automatic differentiation is used here to efficiently compute the partial derivatives that involve complicated relationships between parameters and integrals. This is necessary, for example, when considering the partial derivative of the objective function with respect to the level set function, which involves differentiating through the unfitted integration procedure. Analytical derivatives are not easily found in these cases.
The adjoint method is still required to ensure that the gradient is computed efficiently, i.e. the adjoint equation is solved using partial derivative information from automatic differentiation. That is to say, we use automatic differentiation in conjunction with the adjoint method.

After applying a design update, the level set is reinitialised to a signed distance function. This is done to control the gradients of the level set function and is important for the stability of the evolution. Reinitialisation involves solving the following problem to steady state:
\begin{align}\label{eqn: reinit}
\begin{cases}
    \dfrac{\partial \varphi}{\partial t}(t,\boldsymbol{x}) + \operatorname{sign}(\varphi_0(\boldsymbol{x}))\left(\lvert\boldsymbol{\nabla}\varphi(t,\boldsymbol{x})\rvert-1\right) = 0,\\
    \varphi(0,\boldsymbol{x})=\varphi_0(\boldsymbol{x}),\\
    \boldsymbol{x}\in D, t>0.
\end{cases}
\end{align}
Details on solving this equation can be found in \cite{Wegert2025}.
The methods here are implemented using the Julia \citep{Julia-2017} libraries Gridap \citep{Badia2020,Verdugo2022} and GridapTopOpt \citep{Wegert2025-1}. The reader is referred to \cite{Wegert2025-1} for more details on implementation. The approach is, for the most part, the standard one for level set topology optimisation methods \citep{ALLAIRE20211} and involves only the adaptations required for applicability to the specific CutFEM discretisation.
}

\begin{figure}
	\centering
    \incfig[1]{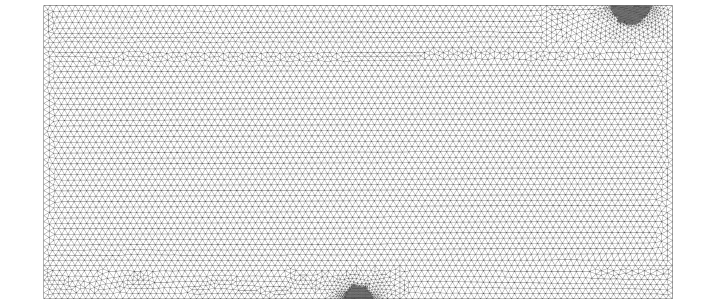}
	\caption{
   \updatedinfirstround{ The unstructured mesh used for the topology optimisation problem that has local refinement around the force input and Dirichlet boundary conditions. The rotation of the computational domain compared to earlier figures is to enable compact presentation. }}
	\label{fig:mesh-independence}
\end{figure}

\updatedinfirstround{
An unstructured mesh is utilised that has local refinement at the points of stress discontinuity around the force input and Dirichlet boundary conditions that are known a priori. The
electrical and mechanical problems are coupled through the stress, so these singularities
can impact the accuracy of the solution for the electrical problem. The utilised mesh with refinement
around these points can be seen in Figure~\ref{fig:mesh-independence}.
The simplectic mesh was generated using Gmsh \citep{Geuzaine2009} with local mesh-size fields, resulting in a mesh composed of 13,706 elements.}

\begin{figure}
	\centering
    \incfig[0.85]{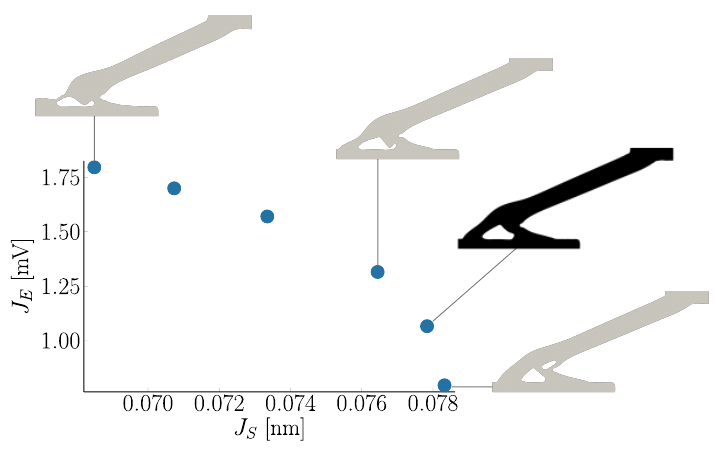}
	\caption{
   \updatedinfirstround{Pareto front of the multi-objective optimisation problem where
   $J_E$ is the electrical objective 
   and 
   $J_S$ is the mechanical objective. 
   \removedinfirstround{The blue points represent the Pareto front solutions obtained by varying $\alpha_E$, the weight of the electrical objective.}
   The geometry in black is the design that we use for the state recovery problem in Section \ref{sec:state-recovery}.}} 
	\label{fig:pareto-front}
\end{figure}

\begin{table}
  \centering
  \captionsetup{labelformat=empty}
  \caption{\updatedinfirstround{\makebox[\linewidth][c]{\textbf{Table~\thetable:} \updatedinfirstround{Structural and electrical performance of candidate geometries along the Pareto front with $ \alpha_S=10^{11}$}}}}
  \label{tab:pareto-front}
  \begin{tabular}{S[round-mode=places,round-precision=0,table-number-alignment=center,table-format=4.0] @{\hspace{1.0em}}S[round-mode=places,round-precision=4,table-number-alignment=center,table-format=0.4] S[round-mode=places,round-precision=3,table-number-alignment=center,table-format=1.3]}
    \toprule
    \multicolumn{1}{c}{\(\alpha_E\)} & \multicolumn{1}{c}{\(J_S\,[\mathrm{nm}]\)} & \multicolumn{1}{c}{\(J_E\,[\mathrm{mV}]\)}\\
    \midrule
    0 & 0.078322849993598315 & 0.7946952301212673\\
    500 & 0.077833808350491099 & 1.0664076195324446\\
    1000 & 0.076445788881225371 & 1.3155337867925554\\
    1500 & 0.073349675906369413 & 1.5706381429471394\\
    2000 & 0.070735888751274512 & 1.700028137121945\\
    3000 & 0.068494584929391797 & 1.7966272067978517\\
    \bottomrule
  \end{tabular}
\end{table}

The optimisation problem is solved by fixing $\alpha_S=10^{11}$ and taking a range of values of $\alpha_E$ to obtain a Pareto front, shown in Figure \ref{fig:pareto-front}. The associated performance values are given in Table \ref{tab:pareto-front}. The large value of $\alpha_S$ is due to the small scale of the mechanical objective that integrates a small displacement over a small area.
\removedinfirstround{The selected design is shown in black in Figure \ref{fig:pareto-front} and corresponds to an electrical objective weight of $\alpha_E = 500$. }
\updatedinfirstround{The structural performance of the gripper is fundamentally important for the gripper to be able to perform quality grasps and our goal for a jointly optimised gripper is for the electrical performance to be improved as much as possible without significantly compromising the gripper's structural performance. We therefore select the design shown in black in Figure \ref{fig:pareto-front} that has an electrical objective weight of $\alpha_E=500$. This design provides a significant improvement in the electrical objective compared to the design with $\alpha_E=0$ without significantly compromising the structural performance. For $\alpha_E$ values greater than 500, the electrical objective is improved further but with a more significant compromise in the structural performance, which is deemed undesirable for our application. The selected design is similar to the one obtained with $\alpha_E=0$, i.e. with only the structural objective active, but with a noticeably thinner diagonal member near the gripper tip. This member, which is important for the structural performance of the gripper, is seen to become too thin at $\alpha_E$ values greater than 500. The selected design is the one adopted to perform the reconstructions in Section~\ref{sec:state-recovery}.}

\begin{figure}
	\centering
    \incfig[1.0]{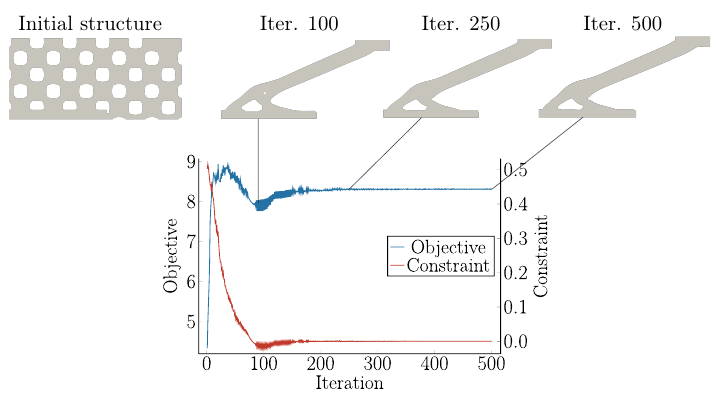}
	\caption{\updatedinfirstround{The iteration history of optimisation objective and volume constraint for the topology optimisation problem with $\alpha_E=500$. The initial structure, some intermediate structures and the final optimised design are also shown.}
   }
	\label{fig:topology-convergence}
\end{figure}

\updatedinfirstround{
Using the unfitted formulation, the boundary of the material domain is represented piecewise linearly within
each finite element. Consequently, even with a relatively coarse mesh, the geometry can be represented
with good accuracy because of the interelement subtriangulation \citep{Andreasen2020,Burman2015}. 
To confirm accuracy of the 
solution to the structural and electrical problems, one level of red-green refinement \citep{Zhao2010}
is applied to the mesh to obtain a mesh with 54,072 elements. The resulting solution fields for the selected topology-optimised design are
compared to those obtained using the original unstructured mesh shown in Figure \ref{fig:mesh-independence}. The relative difference between the two solutions is less than 1\% in the $L^2(\Omega)$
norm for both the electrical and mechanical solution fields, suggesting that the original unstructured
mesh with 13,706 elements is sufficient for simulation of the problem.
 Importantly, the objective and constraint
measurements are accurate with the utilised mesh and can be used to draw conclusions about the optimisation approach and optimised designs.}

\updatedinfirstround{The iteration history of the optimisation objective and volume constraint for the selected design is shown in Figure \ref{fig:topology-convergence}. Oscillations in the objective and constraint can be observed, however the constraint is satisfied at the end of the optimisation and the objective is seen to have converged. 
\updated{One possible source of oscillations during optimisation may be that the objective function considers point-wise measurements of the potential as opposed to average integrated quantities. The oscillations may also be an artifact of the first-order optimisation method used.
We consider that the oscillations are not a significant issue for our purposes, as the objective is seen to converge and the final design is sufficient for our purpose of investigating the potential of the proposed electrical objective to improve reconstruction performance. However, in future work, it may be beneficial to consider a more advanced optimisation method, which could potentially allow for a more optimal design to be obtained.}

The initial geometry, some intermediate structures and the final optimised design are also shown in Figure \ref{fig:topology-convergence}. The initial geometry is a rectangular block with regularly spaced holes. All but one of the holes in the initial design are seen to close as the optimisation progresses. We therefore expect that the final design is not highly sensitive to the choice of initial design. The optimised design obtained is sufficient for our purpose, which is to investigate the potential of the proposed electrical objective to improve reconstruction performance. }

\section{State recovery}\label{sec:state-recovery}

To prevent structural failure and ensure quality grasping, it is desirable for the gripper's control system to have displacement information available.
During operation, however, the location and shape of the contact object is unknown, so the system expressed in Eq. \eqref{eq:weak-formulation} is under-resolved and cannot be solved to obtain the displacement field.
Measurements of the electrical field can be taken to determine the shape of the contact object indirectly. In particular, by solving an inverse problem the shape of the contact object can be determined and the system Eq. \eqref{eq:weak-formulation} can be used to compute the displacement field. 
This step is performed after doing a topology optimisation of the gripper, i.e. $\varphi_{\text{grip}}$ is now fixed to the level set function associated with the black gripper in Figure \ref{fig:pareto-front}. 

We choose to formulate the shape reconstruction inverse problem as follows:
\begin{equation}
  \begin{aligned}
    \underset{s \in V^1 }{\text{min}}  \       \sum_{i \in I_A} 
      \big\lVert \phi_i - \phi_{i,\text{obs}}
    \big\rVert_{\ell^2(I_E)} & \\
    \text{s.t}  \ R_{d}(\bm{u}_{},\bm{v}_{},\varphi,s) + \sum_{j \in J} R_\phi((\bm{u},\phi_{j}),(\bm{v}_{},\eta_{j}),\varphi) 
    = 0, & \quad \forall\ (\bm{v}_{},\eta_{}) \in \mathbf{V}_{0} \times V_{i}, \\
  \end{aligned}
\label{eq:opt-problem}
\end{equation}
where the constraint is the residual for the forward problem as defined in Equation~\eqref{eq:weak-formulation}.
The objective function for this recovery problem is the minimisation of the sum of the $\ell^2$ norms of the difference between the observed electrical potential $\phi_{i,\text{obs}}$ and the approximated electrical potential $\phi_{i}$ at the measurement electrodes for all source-sink patterns. The observed potentials are generated synthetically, firstly without noise, by solving Equation~\eqref{eq:weak-formulation} for a known contact object shape. 
The approximated potentials are obtained by solving Equation~\eqref{eq:weak-formulation} for a guessed contact object shape.
It is expected that the minimiser of this loss function will correspond to a reasonable approximation of the contact object, i.e. the guessed shape will be close to the actual shape.
\updated{We formulate the objective in terms of electrical potentials, rather than displacement or deformation fields, because the electrical potential is the only data available to the sensing system: the gripper has no independent means of directly measuring its own displacement field, which is precisely the information this sensing method is intended to provide. Formulating the objective in terms of the available boundary data rather than the unknown target quantity is standard practice for inverse problems of this type, such as electrical impedance tomography \citep{youssef2024past}.}
In the remainder of this section, we perform multiple experiments using this inverse framework.
The inverse problem is essentially a level set topology optimisation problem where the design variable is the contact object shape $\varphi_{\text{obj}}$ and the objective is to match the observed electrical potentials.

\removedinfirstround{The (unconstrained) Augmented Lagrangian approach is used where the sensitivities are again computed using automatic differentiation.}
Again, the packages Gridap and GridapTopOpt are used for implementation of the problem. 
\updatedinfirstround{A gradient descent method is used to solve the optimisation problem. This method defaults from the augmented Langrangian approach in GridapTopOpt when constraints are not present \citep{Wegert2025-1}. The sensitivities are again all computed using automatic differentiation.}
To aid convergence of the optimiser, we find that it is advantageous to have a smooth conductivity field when solving the electrical problem. To obtain this, we use a $L_2$ projection of the conductivity $\kappa(\bm{u})$ onto the space $C^0$.
\updatedinfirstround{Note that explicit regularisation of the state reconstruction objective function is not included in this formulation. However the optimisation is performed using the level set method described in Section~\ref{sec:topology-optimisation} that involves a Hilbertian extension regularisation of the gradient and a reinitialisation strategy to maintain the signed distance property. These regularisation strategies are found to be sufficient for the convergence of the optimiser.}

\begin{figure}[]
	\centering
    \begin{tabular}{cc}
    \includegraphics[width=0.55\linewidth]{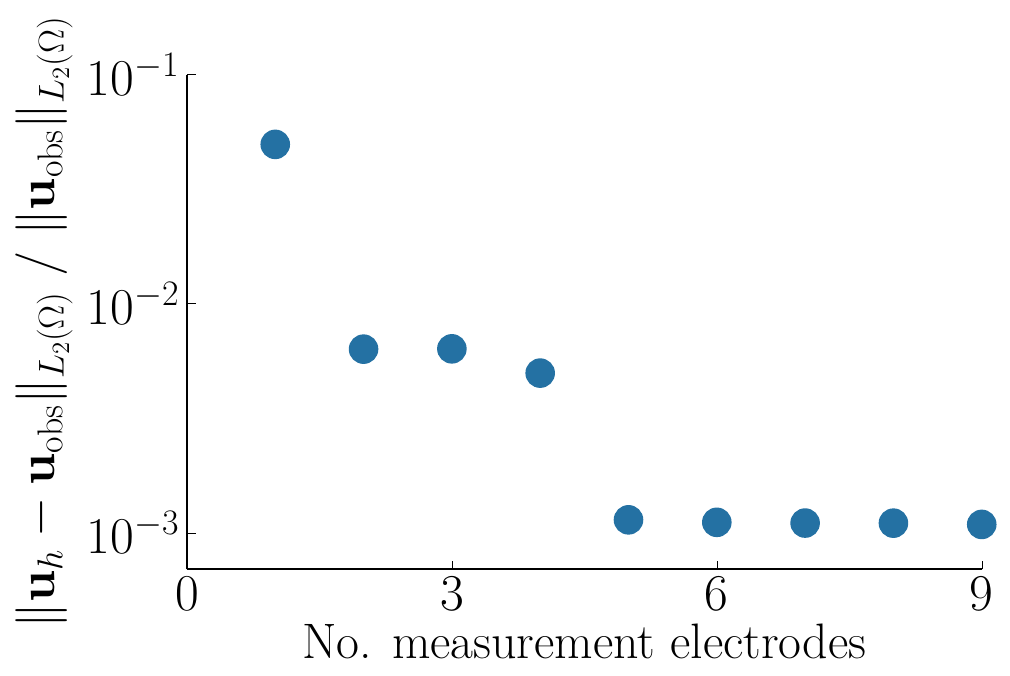} & 
    \includegraphics[width=0.44\linewidth]{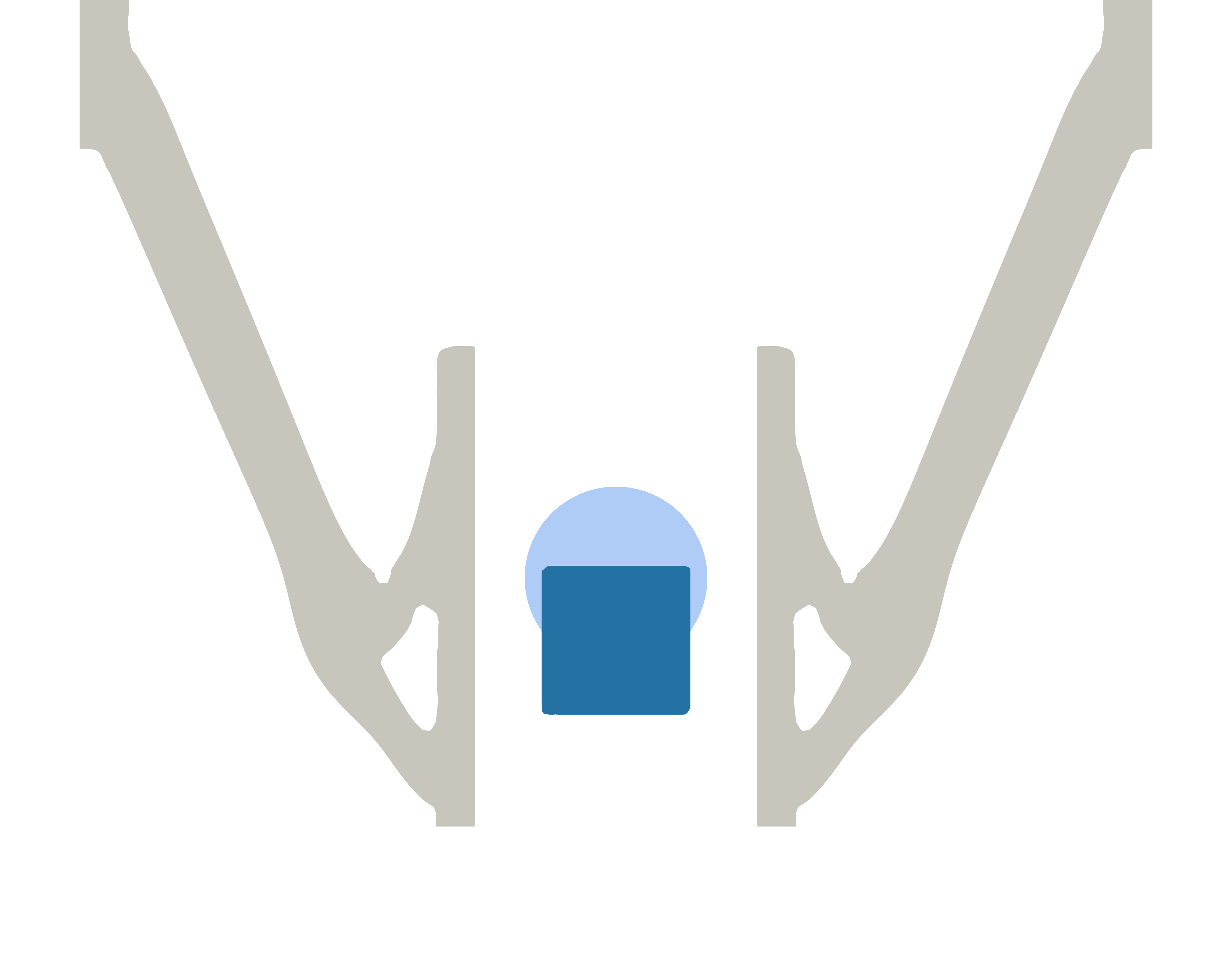} 
    \end{tabular}
	\caption{
    Convergence plot of the relative error in the reconstructed displacement field as the number of measurement electrodes is increased. 
    \removedinfirstround{The number of measurements electrodes $N$ corresponds to using the electrodes numbered from 1 to $N$ in Figure \ref{fig:current-example}.}
    The reconstruction results here are for the recovery when gripping a square object with an initial guess of a circle using the mixed-objective optimised gripper design (right).
        }
	\label{fig:electrode-convergence}
\end{figure}
\begin{figure}[]
    \centering
    \begin{tabular}{cc}

    \includegraphics[width=0.49\textwidth]{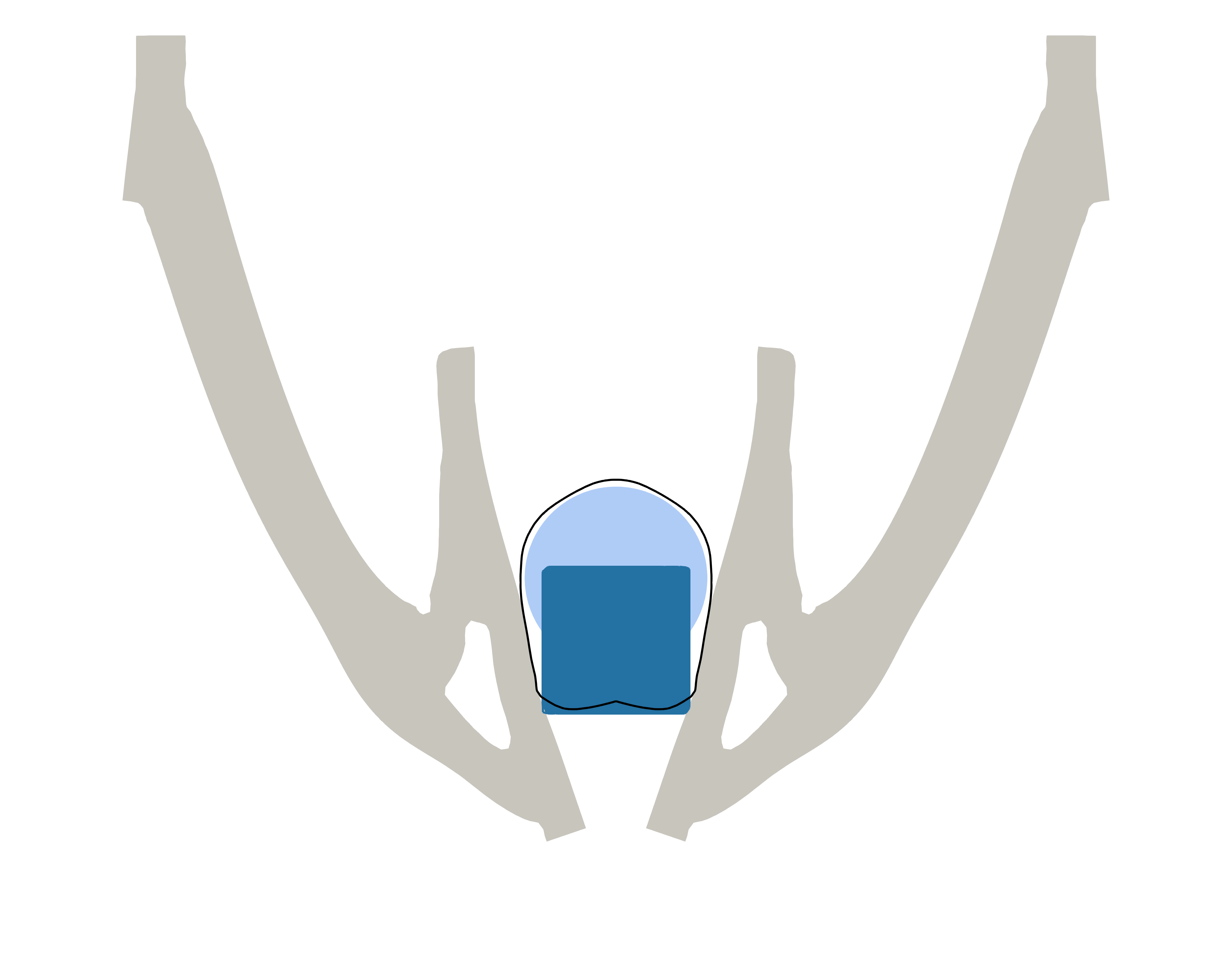} & 
    \includegraphics[width=0.49\textwidth]{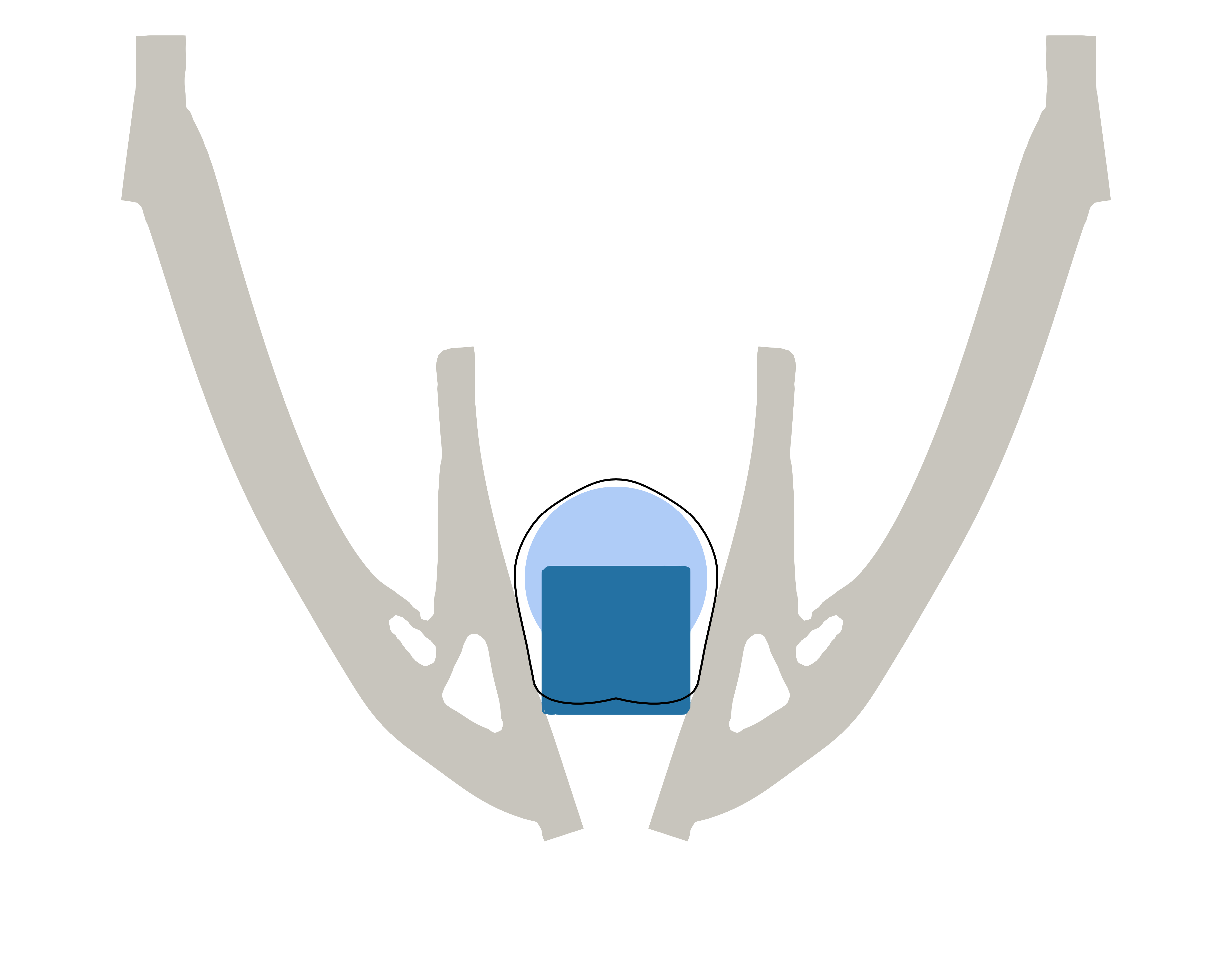} \\
    [-8em] 

    \includegraphics[width=0.49\textwidth]{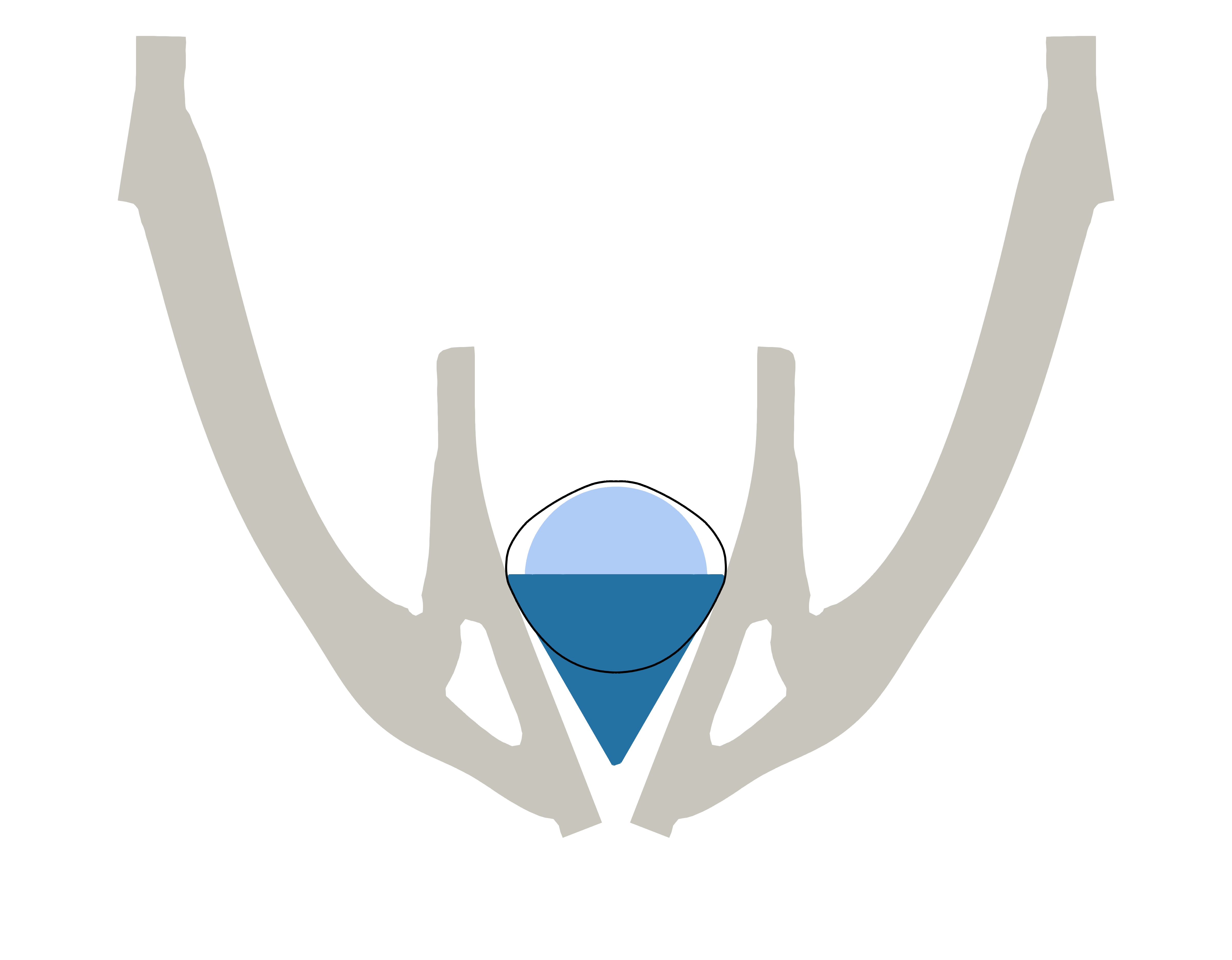} &
    \includegraphics[width=0.49\textwidth]{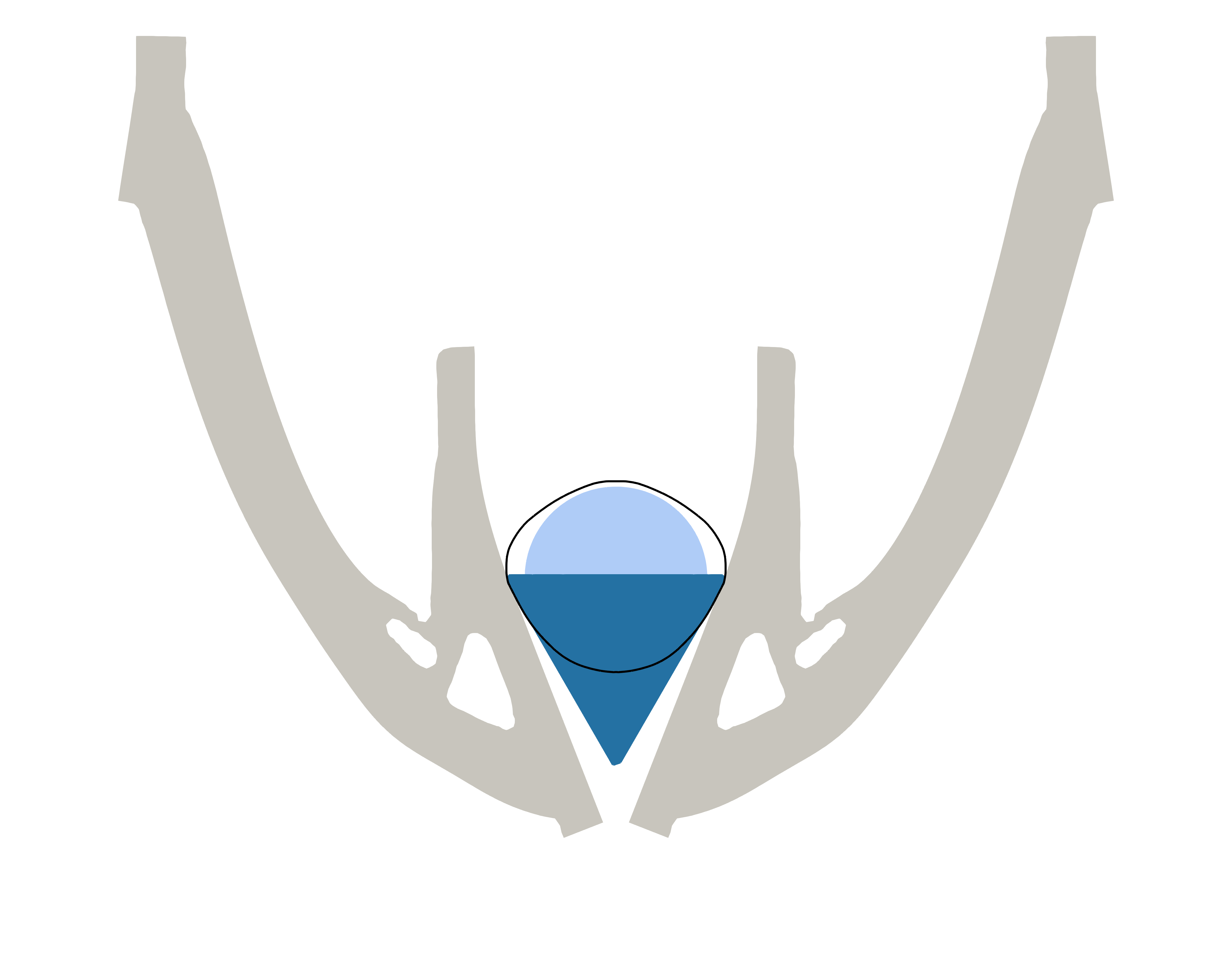} \\
    [-8em] 

    \includegraphics[width=0.49\textwidth]{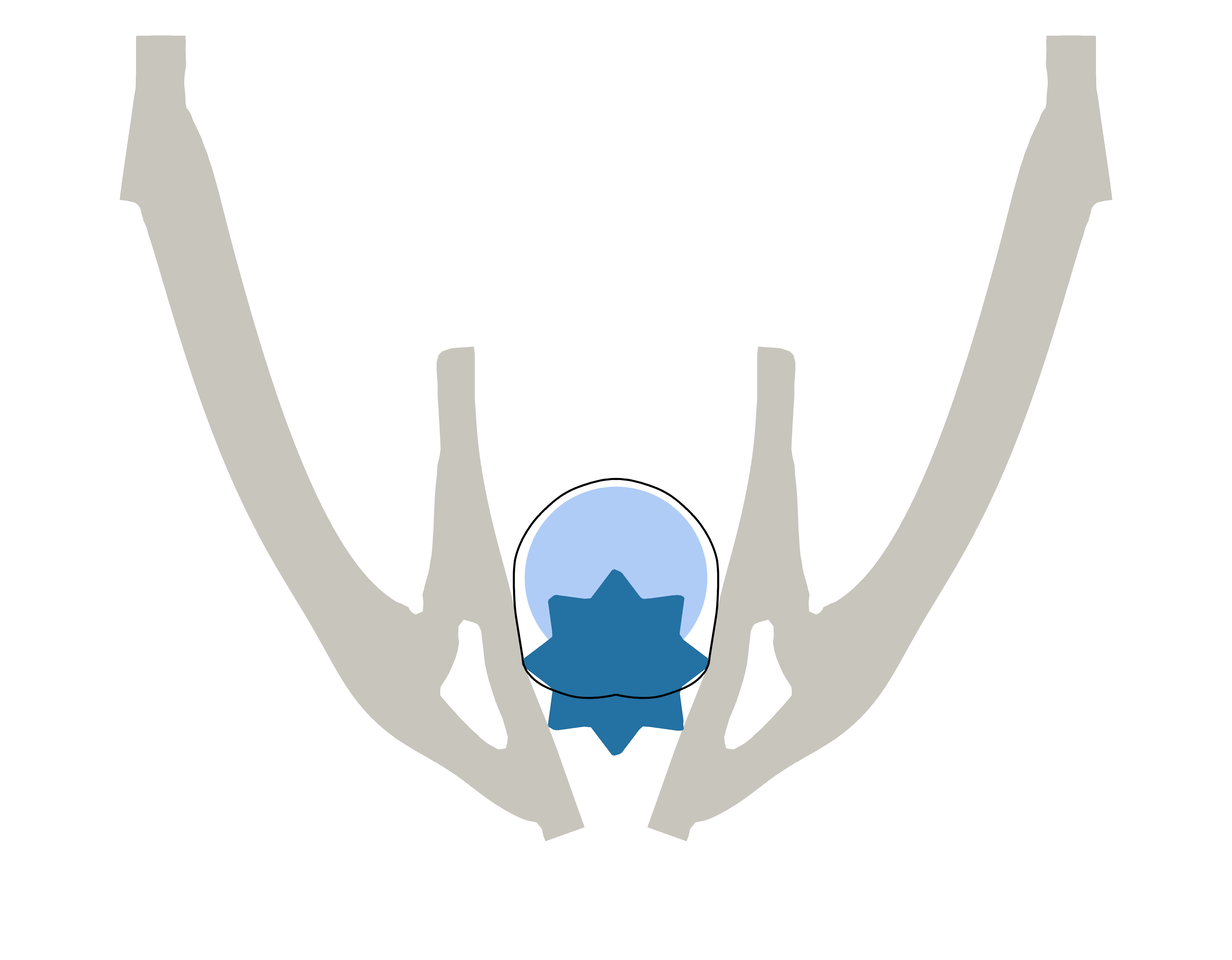} & 
    \includegraphics[width=0.49\textwidth]{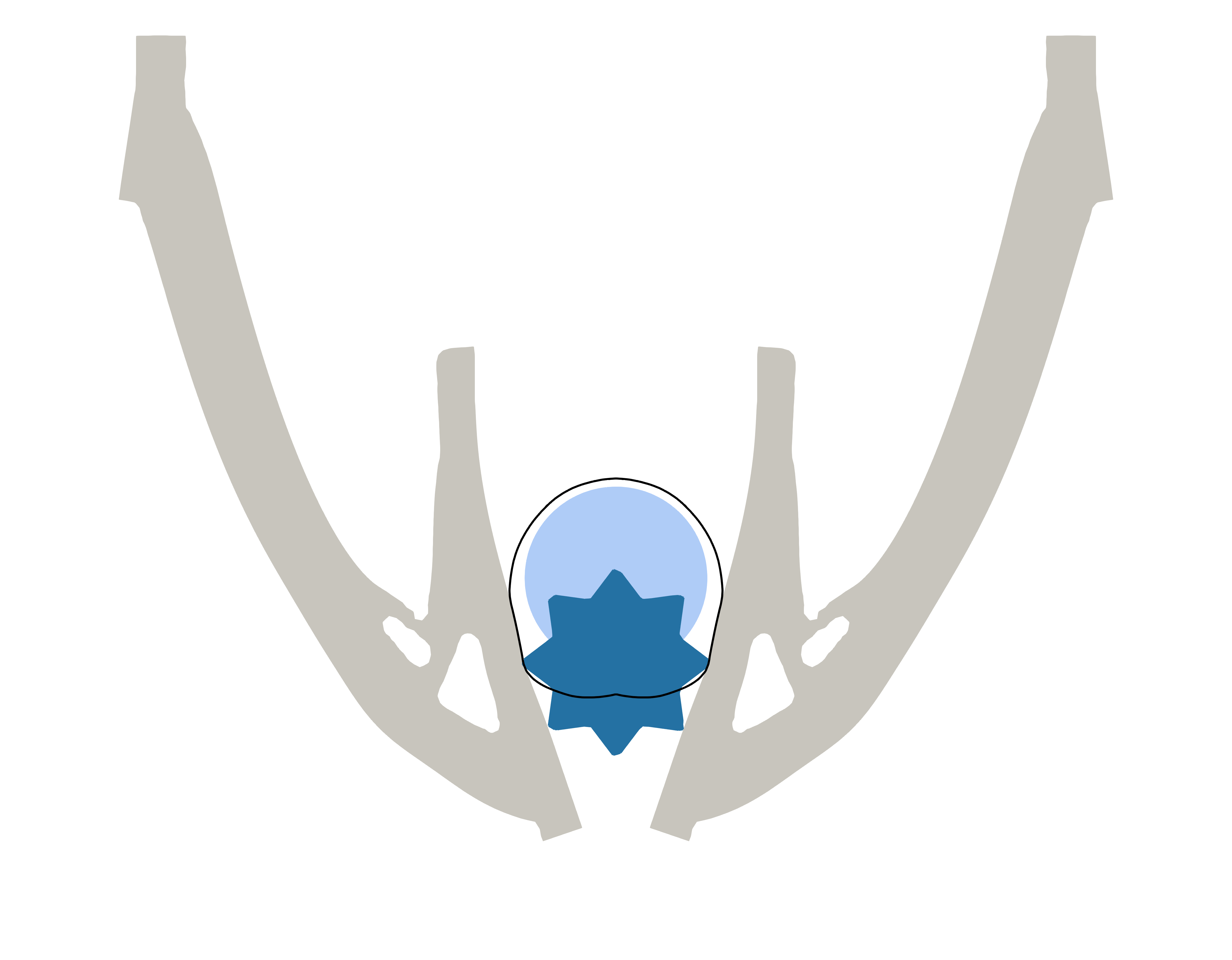} \\
    [-2em] 

    \shortstack{Mixed-objective design} & \shortstack{Mechanical-only design} \\

\end{tabular}
\caption{Shape reconstruction visualisations. Using an initial guess of a circle (light blue), the target square, triangle and star shapes (dark blue) are reconstructed using a gripper designed with the mixed electrical and mechanical objectives (left column) and a gripper designed considering only the mechanical objective (right column). The reconstructed shape is depicted with a black outline and the recovered deformation field for the gripper is shown grey.}
\label{fig:reco-comparison}
\end{figure}

Firstly, we investigate the effect of the number of measurement electrodes on the reconstruction quality. We measure the performance of the gripper by the relative $L^2$ error between the reconstructed displacement field and the actual displacement field, i.e. $\lVert \bm{u} - \bm{u}_{\text{actual}} \rVert_{L^2(\Omega)}/\lVert \bm{u} \rVert_{L^2(\Omega)}$. The measurement electrodes used for a reconstruction
 are the electrodes numbered from 1 to $N$ in Figure \ref{fig:current-example}.
The results are shown in Figure \ref{fig:electrode-convergence} (left) for reconstruction of a square contact object using the initial guess of a circle. It can be seen that increasing the number of measurement electrodes improves the reconstruction quality for up to 6 electrodes, beyond which the error is not seen to significantly improve. Consequently, we use 6 measurement electrodes throughout this work. The recovered square shape and associated displacement field using 6 measurement electrodes can be seen in the top left of Figure \ref{fig:reco-comparison}.

To investigate the viability of the reconstruction framework for general objects, we experiment with the recovery of a few object shapes using both the mixed-objective and mechanical-only designs. We refer to the selected design in black in Figure \ref{fig:pareto-front} as the mixed-objective design, since it considers both the electrical and mechanical objectives in the topology optimisation problem. 
The design obtained by setting $\alpha_E=0$ is optimised for the mechanical objective only, and is referred to here as the mechanical-only design.
The mechanical-only design is seen in the bottom right hand side of Figure \ref{fig:pareto-front}. A visualisation of the reconstruction results for square, triangular and star objects using both the mixed-objective and mechanical-only gripper designs are given in Figure \ref{fig:reco-comparison}.

\begin{figure}[]
    \centering
    \begin{tabular}{cc}

    \includegraphics[width=0.49\textwidth]{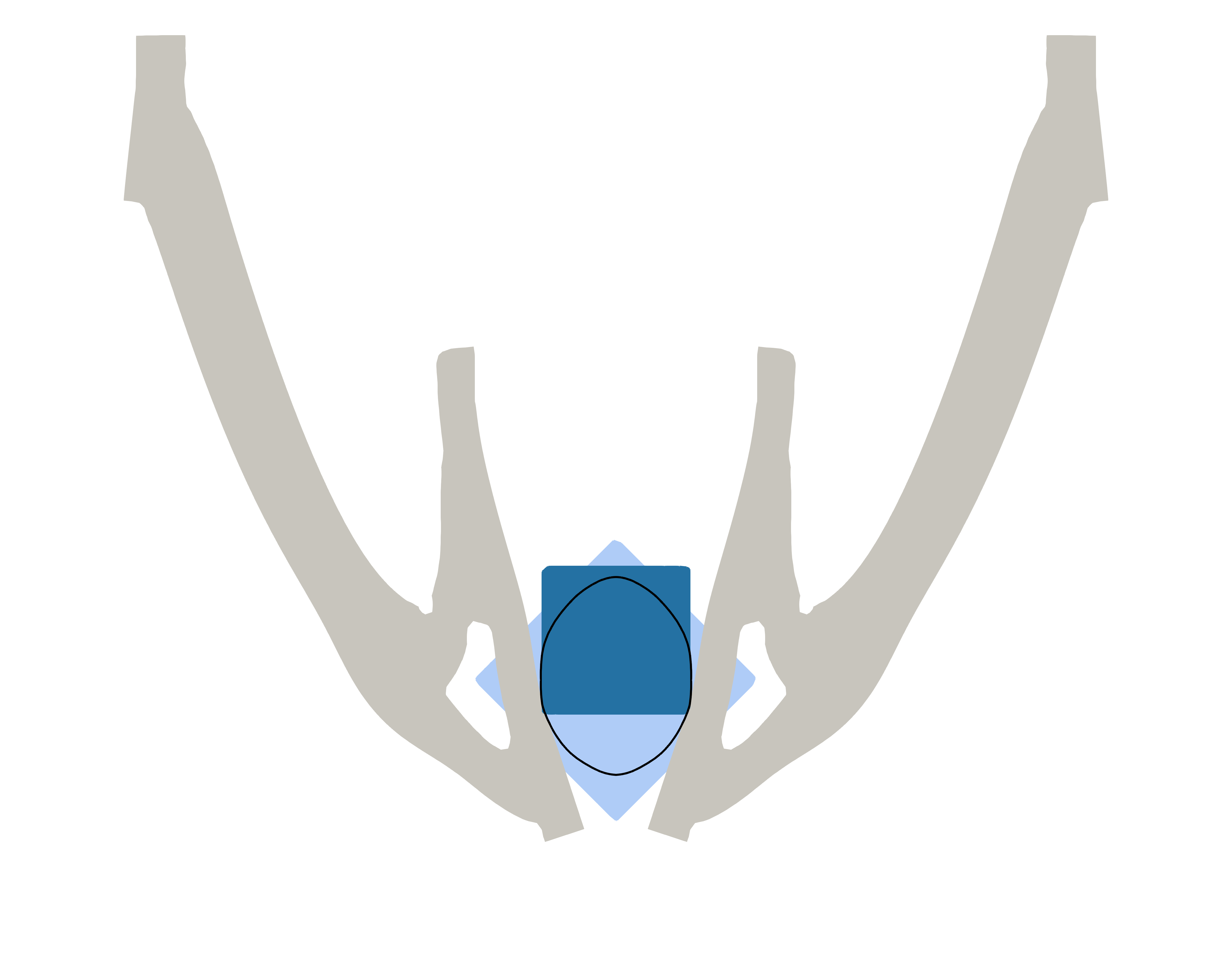} & 
    \includegraphics[width=0.49\textwidth]{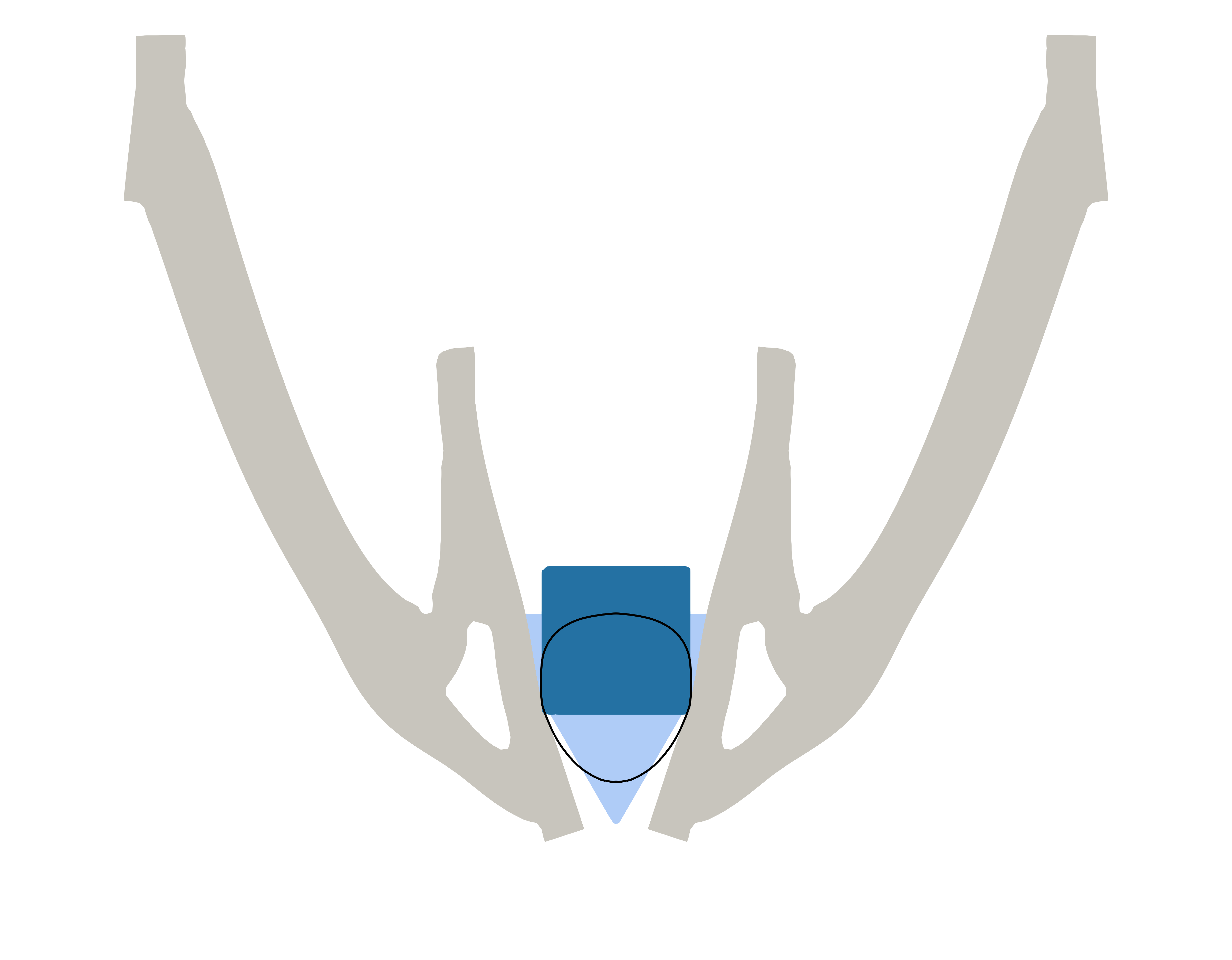} \\
    [-2em] 




\end{tabular}
\caption{Shape reconstruction visualisations using different shapes as the initial guess. A diamond is used as the initial guess on the left hand side and a triangle is used on the right hand side.}
\label{fig:different-init-shapes}
\end{figure}

It can be seen that the contact locations are roughly recovered, with varying degrees of accuracy, for the different shapes using both gripper designs.
It is important to note here that only the contact locations of the gripper on the object are of any importance here, and the geometry of the shapes that are not in contact with the gripper are not relevant.
\updated{This is not a modelling assumption but a direct consequence of the physics of the problem: the geometry of the object outside the contact region does not enter the boundary conditions of the forward problem in Equation~\eqref{eq:d-neumann}, and therefore has no effect on the mechanical or electrical state of the gripper.}
The recovered displacement fields are also used to displace the grippers in Figure \ref{fig:reco-comparison}.
The recovered contact locations and displacement fields are essentially the same with the mixed-objective and the mechanical-only gripper designs for this initial experiment.
These recoveries are however performed without noise corruption, and thus it is expected that both the gripper designs will allow for reasonable reconstruction accuracy.
\updated{As reasonable recoveries are obtained considering various object shapes, we conclude that the number of measurement electrodes is sufficient for the reconstruction. It is possible that a more complex object may require more electrodes, and we consider a rigorous study of the number and placement of electrodes considering all possible object geometries to be an interesting direction of future work.}

\updatedinfirstround{
It is important to note that the state reconstruction problem is inherently ill-posed.  This is clear because the shape of any reconstructed geometry that is not in contact with the gripper is irrelevant, and is also made clear by considering a scenario in which the object is not in contact with the gripper at all. In Figure \ref{fig:different-init-shapes},  we explore sensitivity of the shape reconstruction problem to the initial guess. Figure \ref{fig:different-init-shapes} shows effective recovery of the contact geometry for the square target shape from diamond and triangular initial shapes. However, the shape recovery problem is nonconvex, and shape reconstruction can fail for initial guesses that are very far from the solution. This was seen to occur in the case where the initial shape was located close to the Dirichlet boundary condition but the contact point was close to the gripper tip. It is possible that additional regularisation strategies could improve the robustness of the approach and this exploration has been left to future work. }

\begin{figure}[]
    \centering
    \begin{tabular}{cc}
    \shortstack{} & \shortstack{} \\
    [+6em]
    0\% noise & 0\% noise \\
    [-8em] 
    \includegraphics[width=0.49\textwidth]{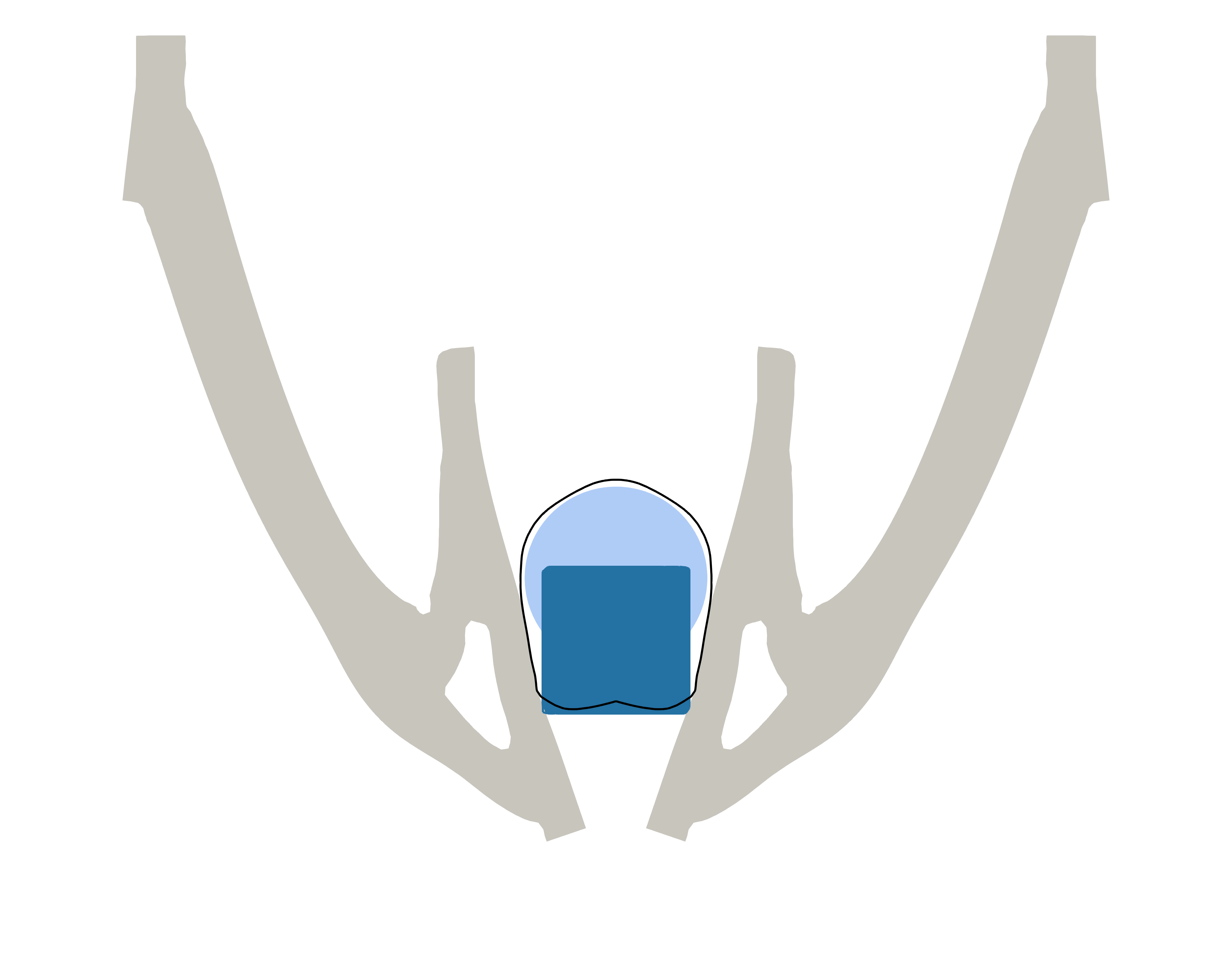} & 
    \includegraphics[width=0.49\textwidth]{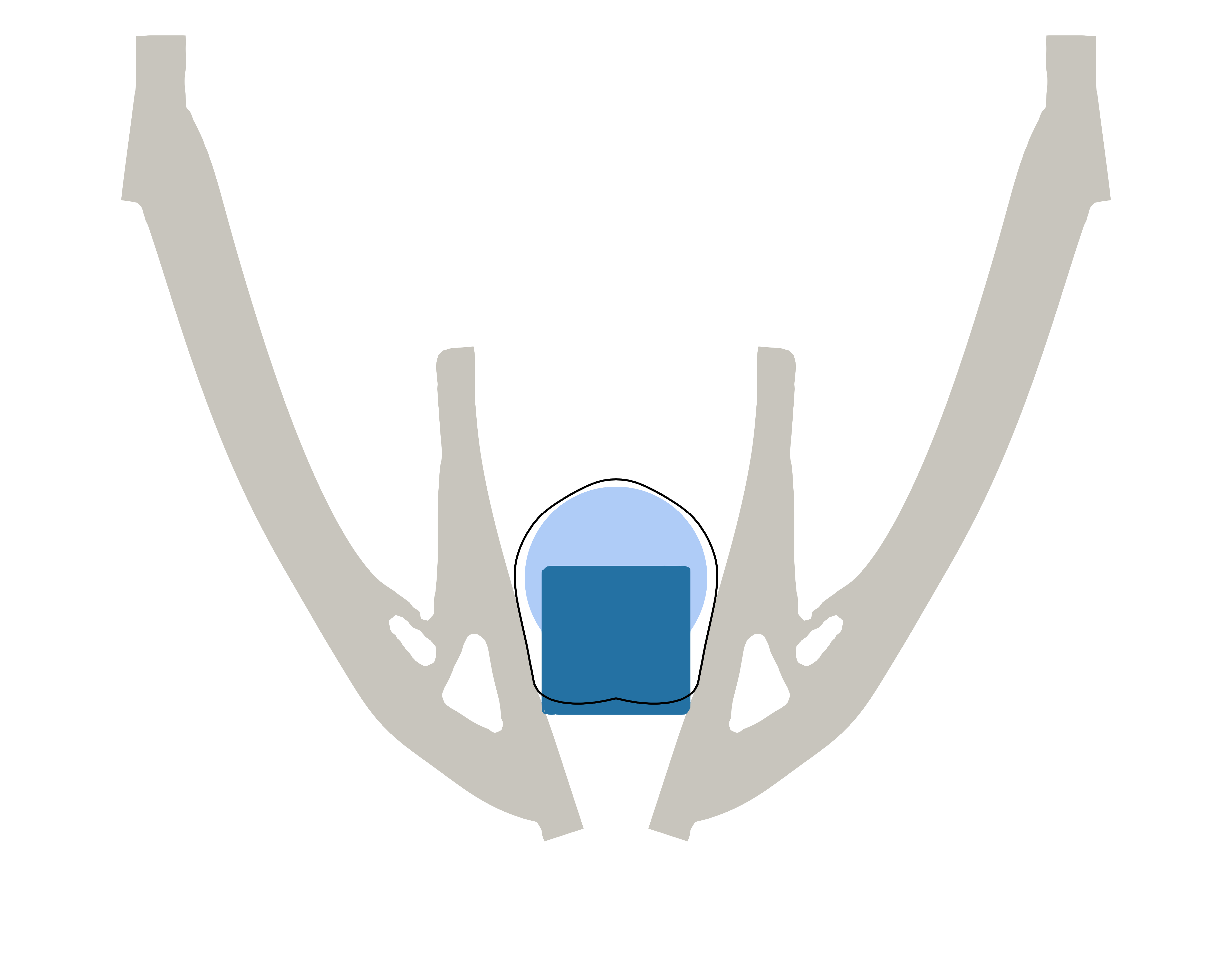} \\
    2.5\% noise & 2.5\% noise \\
    [-8em] 
    \includegraphics[width=0.49\textwidth]{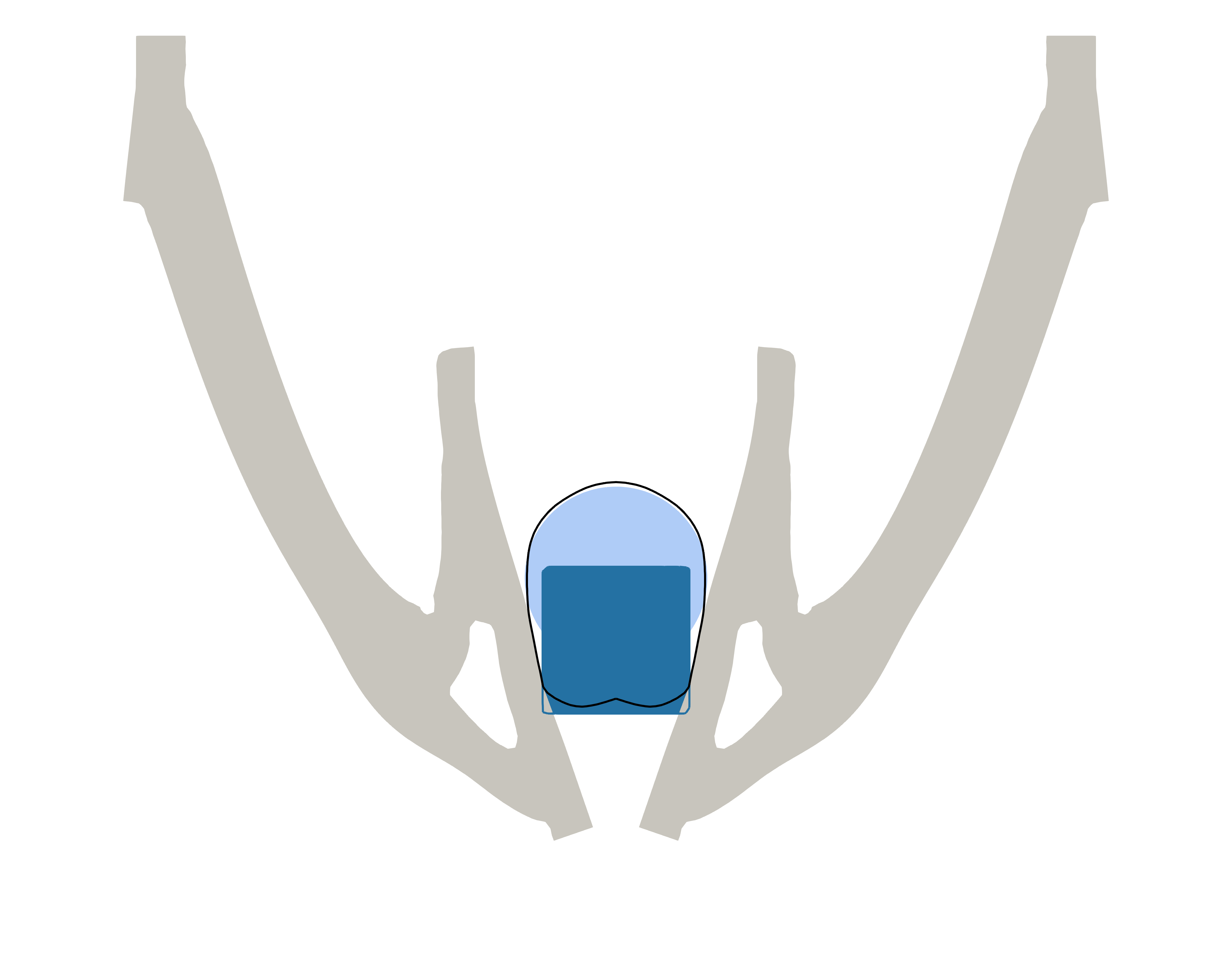} & 
    \includegraphics[width=0.49\textwidth]{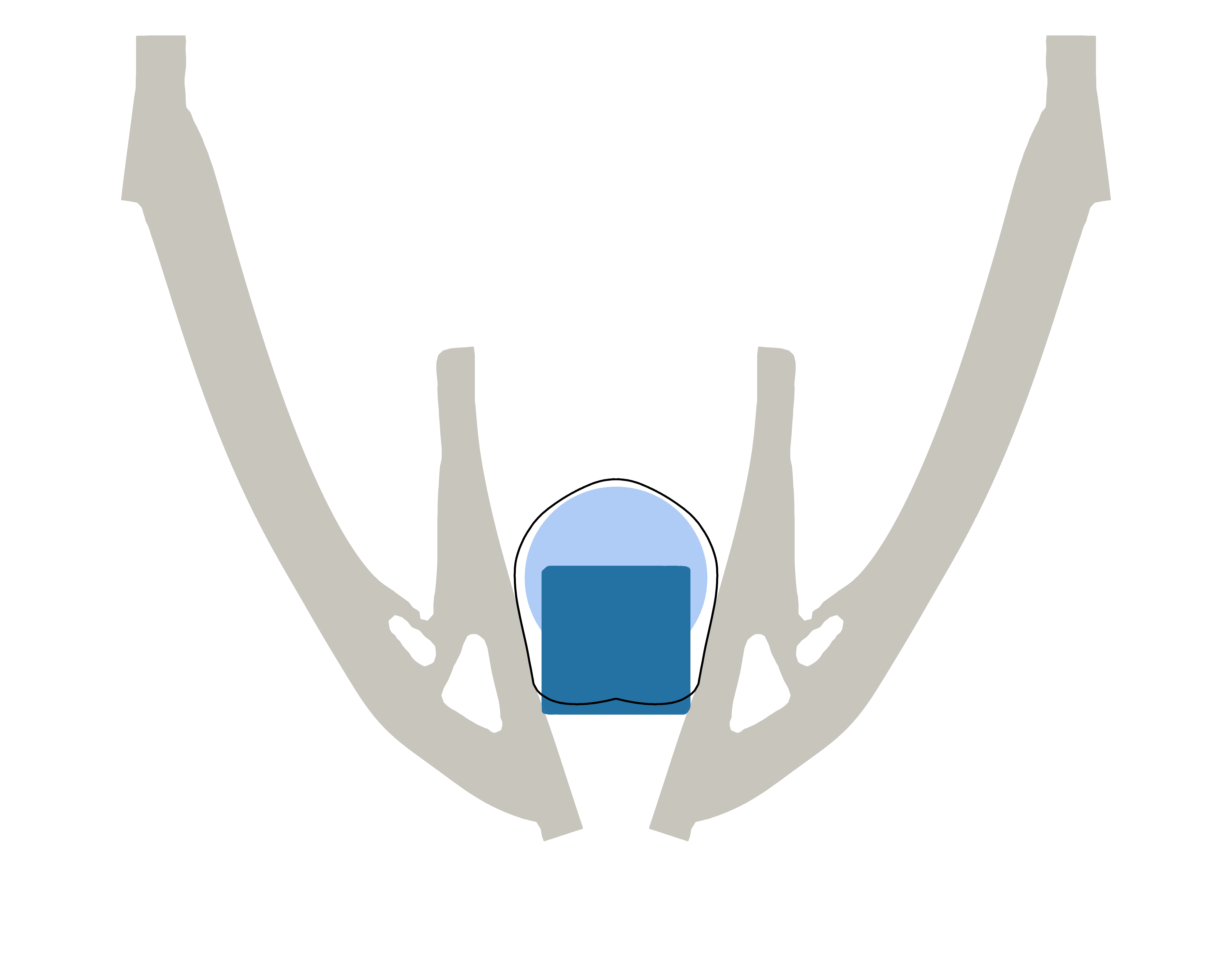} \\
    5\% noise & 5\% noise \\
    [-8em] 
    \includegraphics[width=0.49\textwidth]{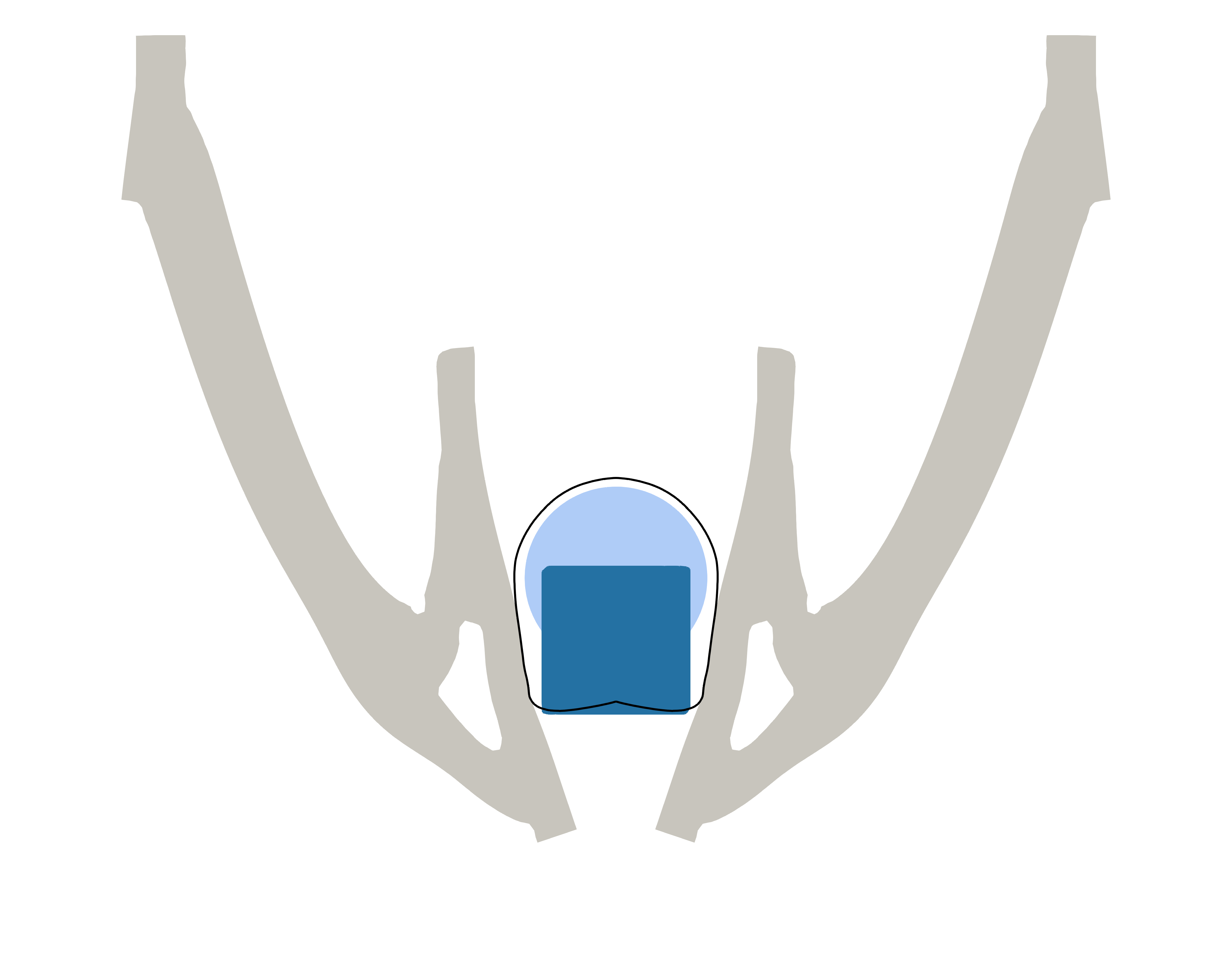} & 
    \includegraphics[width=0.49\textwidth]{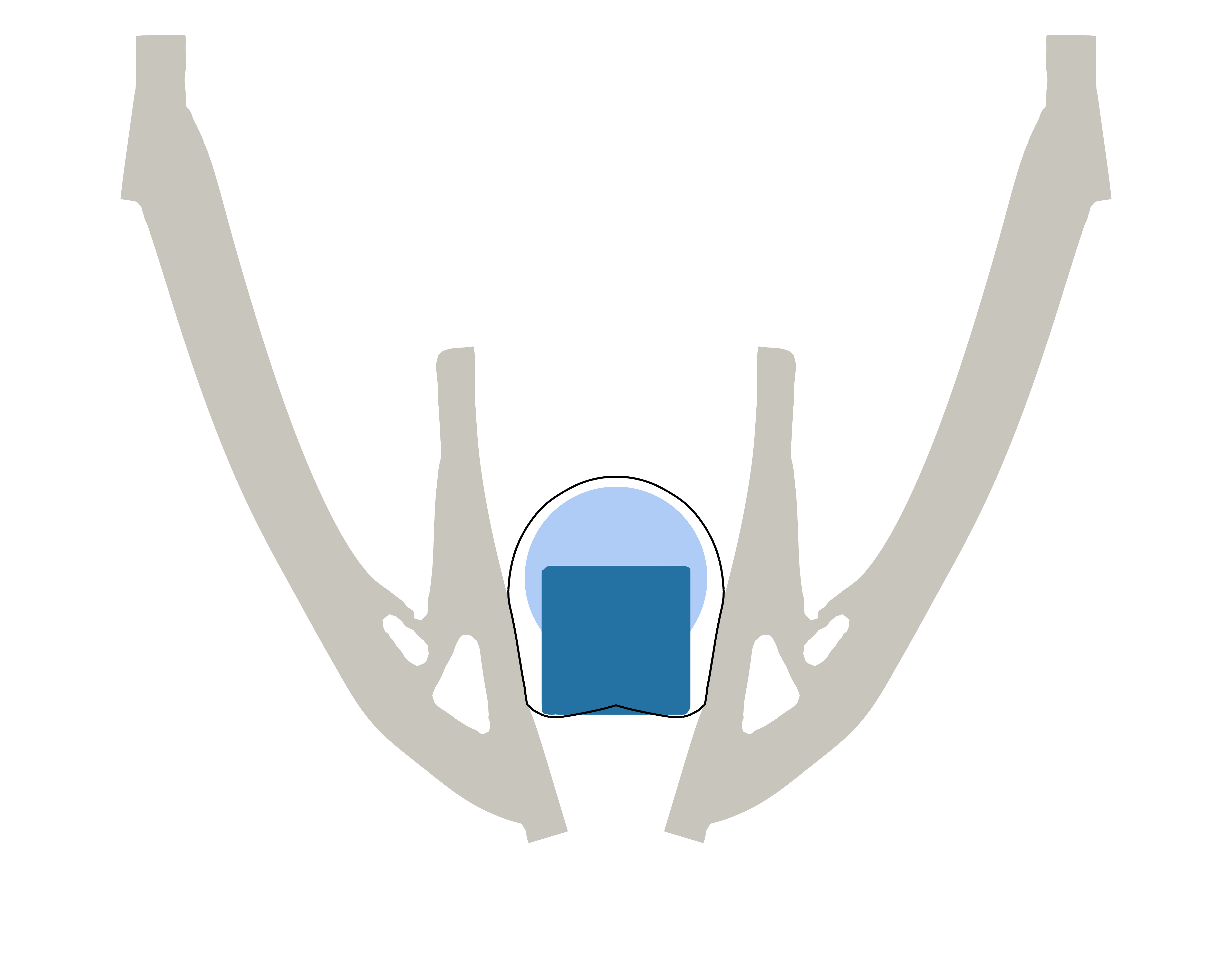} \\
    10\% noise & 10\% noise \\
    [-8em] 
    \includegraphics[width=0.49\textwidth]{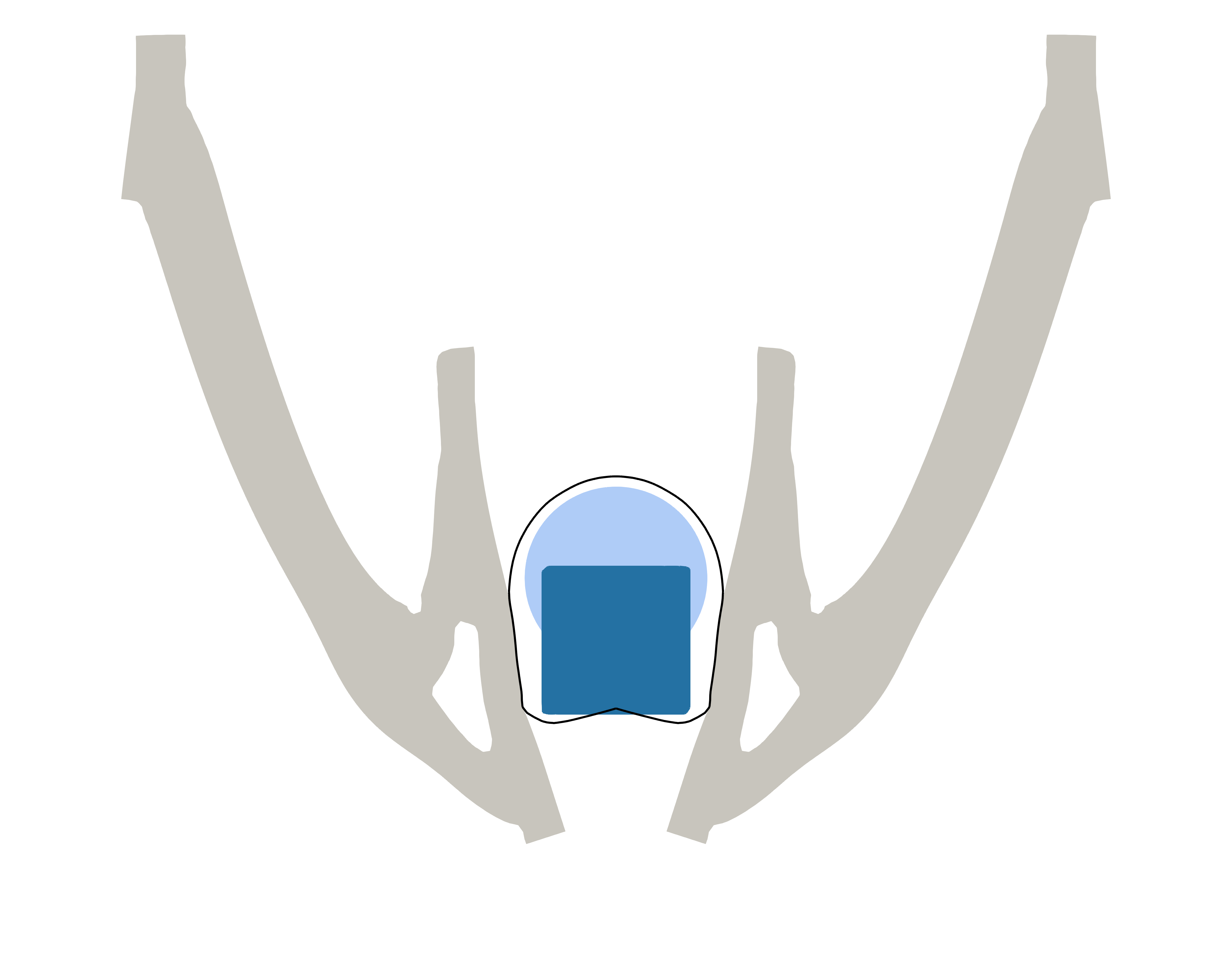} & 
    \includegraphics[width=0.49\textwidth]{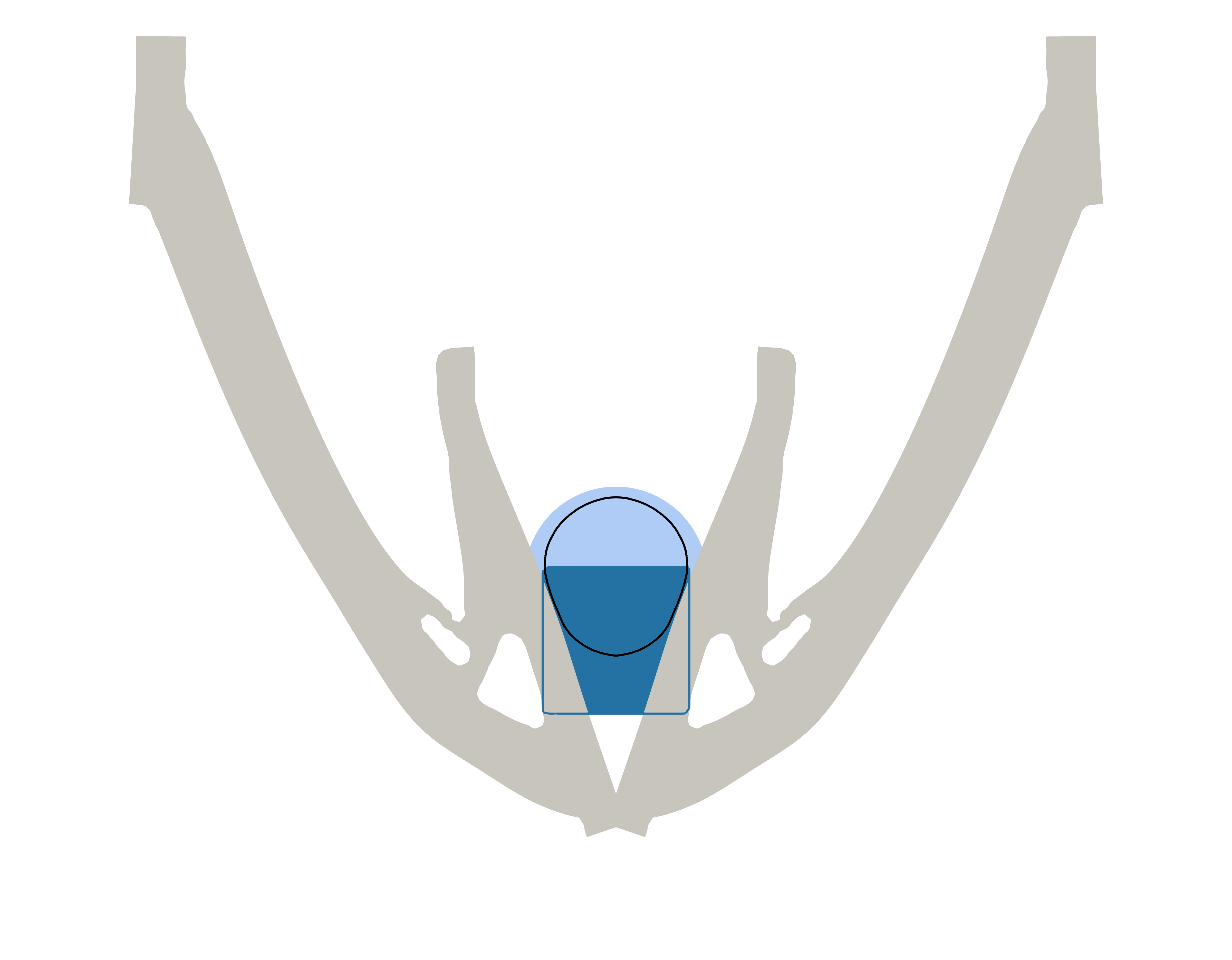} \\
        [-2em] 
    \shortstack{Mixed-objective design} & \shortstack{Mechanical-only design} \\
    \end{tabular}
\caption{Shape reconstruction visualisations with noise corrupted signals. Using an initial guess of a circle (light blue), the target square (dark blue) is reconstructed using a gripper designed with the mixed electrical and mechanical objectives (left column) and a gripper designed considering only the mechanical objective (right column). The reconstructed shape is depicted with a black outline and the recovered deformation field for the gripper is shown in grey. \removedinfirstround{The experiments are conducted with 20 random seeds of noise and the results with a reconstruction error closest to the mean is shown here.}
}
\label{tab:noise-comparison}
\end{figure}

If a larger signal is measured during gripper operation, then the reconstruction may be more robust to a fixed level of noise in the measurement. This is precisely the motivation for the electrical topology optimisation objective used in Section \ref{sec:topology-optimisation}. In the following experiments, we aim to compare the reconstruction robustness between the gripper designed 
when including the electrical objective and the gripper designed when excluding the electrical objective.
We consider here the reconstruction of a square contact object using the initial guess of a circle. The synthetic potential measurements are corrupted with Gaussian noise $\mathcal{N}(0,\sigma^2)$, with a scaled standard deviation $\sigma = \delta \bar{\phi}$ 
where
$\delta$ is the specified noise level and $\bar{\phi}$ is the mean potential reading across all electrodes for all source-sink patterns\removedinfirstround{. The quantity $\Vert \phi_{i,\text{obs}} \Vert_{\ell^2}$ is} computed using the mixed-objective gripper design\removedinfirstround{ and defines the noise scaling for all subsequent tests}. We consider noise levels of $2.5\%$, $5\%$ and $10\%$. For each noise level, we perform 20 reconstructions with different random seeds and report the median reconstruction error and interquartile range.
The reconstruction errors are shown in Table \ref{tab:noisy-shape-reconstruction-results} and the
results with a reconstruction error closest to the mean are shown in Figure \ref{tab:noise-comparison}.

\begin{table}
  \centering
  \caption{Shape reconstruction results with varying levels of noise}
  \label{tab:noisy-shape-reconstruction-results}
  \begin{tabular}{c S[round-precision=3,table-format=1.2e-2,scientific-notation=true] S[round-precision=3,table-format=1.2e-2,scientific-notation=true] S[round-precision=3,table-format=1.2e-2,scientific-notation=true] S[round-precision=3,table-format=1.2e-2,scientific-notation=true]}
    \toprule
    Noise level & \multicolumn{1}{c}{\(\mathrm{median}\)} & \multicolumn{1}{c}{\(\mathrm{IQR}\)} & \multicolumn{1}{c}{\(\mathrm{median}\)} & \multicolumn{1}{c}{\(\mathrm{IQR}\)}\\
    \midrule
    0\% & 0.0011112063121338121 & 0 & 0.0035139467674513598 & 0\\
    2.5\% & 0.0022399674897623624 & 0.0013988650310027436 & 0.0032610542343423704 & 0.0027615071774127467\\
    5\% & 0.0050799080075843144 & 0.010240730868698408 & 0.0090267006921222626 & 0.029741332560868961\\
    10\% & 0.012979029742382481 & 0.0261576109229293 & 0.034284904983214792 & 0.064681737630164798\\
    \bottomrule
  \end{tabular}
\end{table}

It can be seen that the mixed-objective design is more robust to noise than the mechanical-only design, with a smaller median error and interquartile range at each noise level. The mixed-objective design allows for a reasonable reconstruction to be achieved with up to $10\%$ noise corruption. On the other hand, a $10\%$ noise level is seen to greatly deteriorate the reconstruction when using the mechanical-only design.

\begin{figure}[]
	\centering
    \incfig[1]{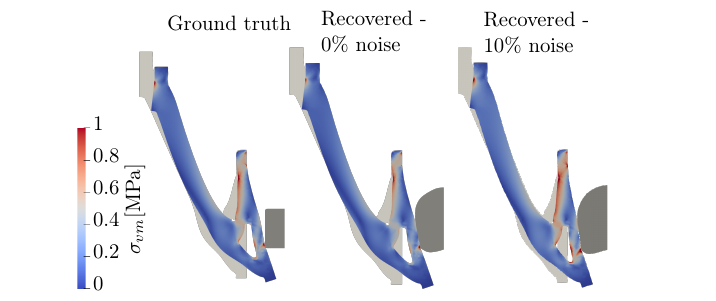}
	\caption{
    Actual displacement and von Mises stress fields (left) and recovered displacement and stress fields without noise (middle) and with 10\% noise (right). 
    The result with a reconstruction error closest to the mean is shown for the noise level of 10\%.
        }
    
	\label{fig:recovered-fields}
\end{figure}

We also plot the actual and recovered displacement and stress fields using the mixed-objective gripper design in Figure \ref{fig:recovered-fields}. It can be seen that the displacement and stress fields are, upon visual inspection, reasonably well recovered using the reconstruction framework and the mixed-objective gripper design with both no noise and 10\% noise.

\updatedinfirstround{We note that the Gaussian noise model was chosen for its simplicity. However, piezoresistive sensors and devices exhibit nonlinear and time-dependent noise including drift and hysteresis. Whilst our method is demonstrably robust to small noise, it cannot compensate for all possible error sources. In practical usage, both noise compensation and periodic recalibration would be required, as is typical for many sensors.}




\section{Discussion}\label{sec:discussion}

\updated{We reiterate that this work is a theoretical and computational proof-of-concept for the proposed piezoresistive sensing and reconstruction method. To the best of our knowledge, no prior work has attempted to recover the full displacement field of a gripper from sparse electrical measurements, and there is consequently no established experimental or computational precedent to benchmark against. Given this, our priority in this first study is to establish, computationally, whether the proposed sensing and reconstruction approach is feasible at all under simplified, idealised conditions, before moving into the experimental phase. In this section, we discuss the simplifying assumptions made in this work and the limitations of the current approach.}


\updatedinfirstround{To demonstrate the concept of the proposed framework, we have assumed small strains and have only considered the case of linear elasticity. The direct application of this approach is therefore only for grippers composed of relatively stiff materials that do not undergo large deformations. 
This is the case, for example, if one were to print the gripper using the commercially available piezoresistive TPU filament NinjaTek Eel \citep{NinjaTek2018Eel} and conduct grasps that involve only the small strains present in the computational experiments in Section \ref{sec:state-recovery}. 
We consider that the general approach may be useful as a first step towards handling scenarios that necessitate nonlinear mechanics, for example in the case of flexible elastomers that undergo large deformations and require real-time control.}

To validate the small strain assumption and the linear elasticity model, we perform an experimental grasp test of a square object using the optimised gripper design printed using the NinjaTek Eel filament. 
\updatedinfirstround{The gripper is \SI{150}{\milli\meter} high, \SI{70}{\milli\meter} wide and \SI{20}{\milli\meter} deep and was printed on a Prusa-XL 3D printer with 100\% infill density, \SI{0.2}{\milli\meter} layer height, \SI{40}{\milli\meter\per\second} print speed, nozzle temperature of \SI{240}{\degree} and bed temperature of \SI{50}{\degree}. 
The tests were performed on an in-house linear testing rig made
by CSIRO, which includes a linear rail, linear pneumatic actuator, and an RGB camera and computer vision system. \updated{The linear rail allows the gripper to be raised and lowered over the stationary object, while the pneumatic actuator provides the downward force required to close the gripper. }
The computer vision system uses classical edge detection and colour thresholding to track the gripper edge and red keypoints during grasping. \updated{As each key point is known to be circular, their centres can be tracked using standard circle fitting after camera calibration (intrinsics and extrinsics), which was undertaken using calibration markers.} In this test, the gripper is lowered over
the object before a pneumatic cylinder closes the gripper by applying a downwards force to the top
left corner. 
This is equivalent to the simulation, but with a \SI{45}{\degree} rotation of the reference frame.}
\updated{For more details on the exact experimental setup, the reader is referred to \cite{Enkera2026}. The difference to the experimental setup in \cite{Enkera2026} is only the specific gripper design and grasped object.}

\begin{figure}[!htbp]
	\centering
    \incfig[1]{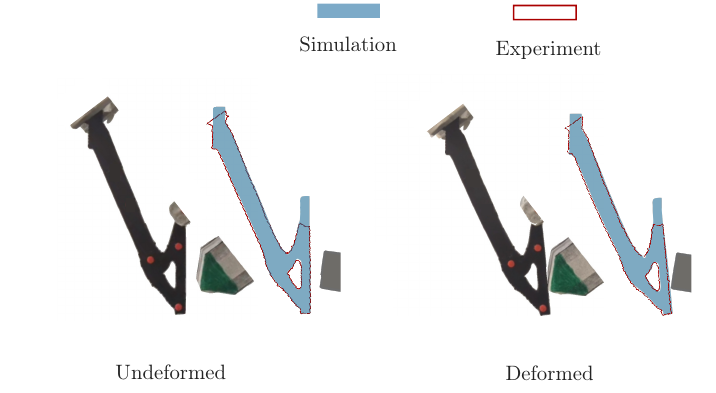}
	\caption{
    Comparison of experimental and simulated displacements. For both the undeformed and deformed shapes, the left image is from experiment and the right image (blue) is from simulation. The gripper boundary obtained in the experiment is overlaid in red on the simulated deformed shape to show agreement. \removedinfirstround{The red points in the experiment are used for location tracking and are not relevant for this comparison.}
    }
	\label{fig:experiment}
\end{figure}

\begin{figure}[!htbp]
	\centering
    \incfig[1]{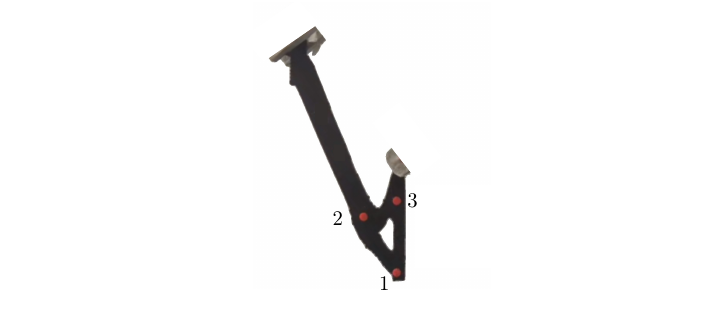}
	\caption{
    The locations of the optical tracking points 1,2 and 3 shown in red in the undeformed configuration.}
	\label{fig:experiment-labels}
\end{figure}

\updatedinfirstround{
The experimental and simulated deformations are compared qualitatively in Figure \ref{fig:experiment}, where the experimental images have been rotated to match the simulation geometry. Visual inspection shows close agreement. A quantitative comparison is enabled by the red keypoints that are optically tracked in the experiment and labelled as shown in Figure \ref{fig:experiment-labels}. Keypoint 3 is not included in the quantitative comparison and is instead used for calibration of the experimental deformed state to account for translation of the supports during the experiment. Keypoint 3 is selected as the calibration point because it is located close to the Dirichlet boundary condition and undergoes a very small deformation. Table~\ref{tab:keypoint_comparison} provides details of the quantitative comparison between the experimentally measured and simulated displacements for keypoints 1 and 2. We note that the experimental images have been rotated into the frame of the simulation prior to these calculations, and the location of the key points is specified relative to the bottom right-hand corner of the computational domain. The relative difference between the simulated and experimental displacement vectors for each point is approximately 2\%, showing excellent agreement between the simulation and experiment with the linear elastic constitutive model.
} 

\updatedinfirstround{We note that agreement between the model and experiment is not expected at large displacements due to the limitations of the linear elasticity model used in the simulations. However, the excellent agreement between the simulated and experimentally measured displacements above suggests that the structural formulation, including the contact model, is reasonable for modelling the deformation of the gripper when grasping similar objects that do not result in large deformations.} \updated{The current framework is also only intended for objects that, themselves, do not undergo significant deformation. An extension of the work to handle objects that undergo large deformation is left to future work. As such, the current framework is not intended for all possible unknown objects, but rather for a restricted class of sufficiently rigid objects.}

\updated{Another important assumption made is the piezoresistive constitutive model. We have assumed a linear relationship between the stress and the electrical resistivity of the material. Experimental characterisations of printable piezoresistive filaments reported in the literature \citep{Georgopoulou2022,Georgopoulou2021,Georgopoulou2020,Riddervold2024} establish a well-defined coupling between mechanical loading and electrical resistance. Among the filaments previously characterised, some exhibit a linear resistance-strain response over the strain ranges considered, while others do not. We adopt a linear constitutive model as representative of this former class, without fitting its parameters to any specific reported characterisation. A baseline experimental characterisation of the resistance or voltage response of a printed gripper under controlled loading, and the extension of the reconstruction framework to the resulting non-linear and history-dependent electrical behaviour, is an important next step and is left to future work. Another improvement to the }\updated{electrical }\updatedinfirstround{model that could be considered is the inclusion of an experimentally derived contact impedance for the electrode-material interface.}
\updated{Due to the simplifying assumptions made in this work and lack of full experimental verification, the results presented here are not expected to be directly applicable to a real-world setting. However, we consider that the proposed framework is a useful first step towards the development of a gripper for which the deformation state can be inferred from electrical measurements.}

There are also approximations related to the numerical method. Using a refined mesh with the current model is likely to improve the reconstruction quality, however, we consider the recovered deformation states to be sufficiently accurate at the current level of refinement for the purpose of this work. Similarly, we could also consider using a larger number of Gauss points for integration in the contact region. 

\updatedinfirstround{
Another important point is that the computational cost of the reconstruction algorithm is currently too high for real-time applications. 
An evaluation of the forward and adjoint problems for the shape reconstruction takes approximately 10 seconds on a single CPU core. We currently use a conservatively small step size and run the algorithm for 1,000 iterations, resulting in a total runtime of approximately 10,000 seconds. We note that this could be significantly improved by utilising a more efficient optimisation algorithm, such as a quasi-Newton method, and by using a parallel implementation of the forward and adjoint solvers. The computational cost would be expected to grow with the number of elements in the computational mesh. Changing the number of measurement electrodes would have a much smaller effect on the computational cost because adding more electrodes increases the number of objective terms in the reconstruction but does not change the number of degrees of freedom in the finite element solves. }

\updatedinfirstround{
To translate the proposed framework to a real-time setting, we would likely need to consider the use of surrogate models to enable fast reconstructions. For example, a neural network could be trained to learn the mapping from electrical measurements to deformation states, using data generated from the inverse problem. This would allow for much faster inference of the deformation state during operation, potentially enabling real-time control. Operator learning for inverse problems is an active area of research that shows promise for this application, where large offline training costs could be justified by the much faster online inference \citep{Molinaro2023}.} \updated{We consider that this constitutes a critical next step for the proposed framework and that the framework in its current form cannot be directly applied to real-time settings.}

\updatedinfirstround{Finally, an aspect that we have not investigated is the optimisation of the sensor placement. The current formulation includes a fixed layout of electrodes, which is a design choice based on practical constraints. Considering these fixed locations, we only investigate how many electrical measurements are required for effective reconstruction.
Optimising the positions of the source-sink and measurement electrodes would likely yield higher-quality reconstructions with fewer electrodes and is considered as an important future extension of this work. The co-design of the sensor placement and geometry would also be an interesting direction for future work, although this would lead to a significant increase in complexity of the computational optimisation problem.}

\begin{table}[h!]
   \caption{\updatedinfirstround{A comparison of the experimentally measured and simulated displacements for keypoints 1 and 2 as shown in Figure~\ref{fig:experiment-labels}.}}
\centering
\begin{tabular}{lll}
\hline
Quantity & Keypoint 1 & Keypoint 2 \\
\hline
undeformed location (mm) & $(-4.38,\ 4.36)$ & $(-24.96,\ 38.83)$ \\
measured displacement, $\boldsymbol{u}_{\mathrm{exp}}$ (mm) & $(7.74, -0.18)$ & $(4.21, -2.60)$ \\
simulated displacement, $\boldsymbol{u}_{\mathrm{sim}}$ (mm) & $(7.77, -0.31)$ & $(4.15, -2.53)$ \\
error, $\|\boldsymbol{u}_{\mathrm{sim}} - \boldsymbol{u}_{\mathrm{exp}} \|_2$ (mm) & 0.14 & 0.10 \\
relative error, $\|\boldsymbol{u}_{\mathrm{sim}} - \boldsymbol{u}_{\mathrm{exp}} \|_2/\|\boldsymbol{u}_{\mathrm{exp}}\|_2$ (\%) & 1.79 & 2.00 \\
\hline
\end{tabular}

\label{tab:keypoint_comparison}
\end{table}

\section{Conclusion}\label{sec:conclusion}
This work introduces a theoretical demonstration of a new class of self-sensing \removedinfirstround{soft }piezoresistive grippers that can infer approximate deformation fields from sparse electrical measurements taken at conveniently placed electrodes. \removed{This \removedinfirstround{enables}\updatedinfirstround{is a \updated{first} step towards} closed-loop control using key quantities such as maximum internal stress.} To design the gripper geometry,
we first developed a topology optimisation formulation to improve the electrical sensitivity while maintaining mechanical performance. To avoid artifacts arising from interpolated materials, we adopt an unfitted finite element method and a level set topology optimisation approach.
Secondly, an inverse reconstruction framework was designed to recover object shapes and gripper displacements for in-operation grippers contacting unknown objects. 
Using the inverse framework, we verify the effectiveness of an electrically-optimised gripper design, showing improved reconstruction robustness based on noisy synthetic electrical observations compared to a gripper designed without the electrical objective. Finally, a physical experiment was conducted to verify the effectiveness of the mechanical model. Optimising the positions of the source-sink and measurement electrodes would likely yield higher-quality reconstructions with fewer electrodes and could be considered as a future extension. Furthermore, an important next step would be to experimentally validate the electrical model and the object shape reconstruction framework \updated{to establish practical applicability}. This will likely be the subject of our future work.

\backmatter


\bmhead{Supplementary information}

Not applicable

\bmhead{Acknowledgements}

\removedinfirstround{Not applicable}\updatedinfirstround{We thank the anonymous reviewers for their constructive feedback that has resulted in significant improvements to the manuscript. }

\section*{Declarations}

\bmhead{Funding}

This work was supported by the Australian Research Council through the Discovery
Projects grant scheme (DP220102759). This research used computational resources provided by:
Queensland University of Technology and the National Computational Infrastructure (NCI) Australia

\bmhead{Conflict of interest/Competing interests}
The authors have no competing interests to declare that are relevant to the content of this
article.

\bmhead{Ethics approval and consent to participate}
Not applicable

\bmhead{Consent for publication}
Consent to publish has been obtained from all authors

\bmhead{Replication of results}
The code to reproduce the results in this paper is available at \url{https://github.com/ConnorMallon/PiezoResistiveGripperDesign}.

\bmhead{Author contributions}
\textit{Connor N. Mallon}: Writing - original draft, Writing - review and editing, Conceptualization, Data curation, Formal analysis, Investigation, Methodology, Software, Validation, Visualization.
\textit{Zachary J Wegert}: Writing - review and editing, Software.
\textit{Anthony P Roberts}: Supervision, Writing - review and editing, Conceptualization, Funding acquisition.
\textit{Joshua Pinskier}: Writing - review and editing, Conceptualization, Supervision. 
\textit{Harry Bowman}: Physical experiments, Writing - review and editing, Conceptualization.
\textit{Vivien J Challis}: Supervision, Writing - review and editing, Conceptualization, Project administration, Funding acquisition, Resources.

\noindent

\begin{appendices}






\end{appendices}




\bibliography{sn-bibliography}

\end{document}